\documentclass[journal]{IEEEtran}

\usepackage[utf8]{inputenc} 
\usepackage[T1]{fontenc}
\usepackage{url}
\usepackage{ifthen}
\usepackage{cite}
\usepackage[cmex10]{amsmath}

\usepackage{graphicx}
\usepackage{amsfonts}
\usepackage{enumitem}
\usepackage{amssymb}
\usepackage{mathrsfs}
\usepackage{bm}
\usepackage{mdwmath}
\usepackage{mdwtab}
\usepackage{dsfont}
\usepackage[ruled,vlined,linesnumbered]{algorithm2e}

\newtheorem{remark}{Remark}
\newtheorem{theorem}{Theorem}
\newtheorem{lemma}{Lemma}

\newtheorem{defi}{Definition}
\newtheorem{coro}{Corollary}

\begin{document}
    \title{Timely Status Update in Massive IoT Systems: Decentralized Scheduling for Wireless Uplinks}
    \author{Zhiyuan Jiang, Bhaskar Krishnamachari, Xi Zheng, Sheng Zhou, Zhisheng Niu,~\IEEEmembership{Fellow,~IEEE}
	\thanks{
    Z. Jiang, X. Zheng, S. Zhou and Z. Niu are with Tsinghua National Laboratory for Information Science and Technology, Tsinghua University, Beijing 100084, China. Emails: \{zhiyuan@, zhengx14@mails., sheng.zhou@, niuzhs@\}tsinghua.edu.cn. 
    
    B. Krishnamachari is with the Ming Hsieh Department of Electrical Engineering, University of Southern California, Los Angeles, CA 90089, USA. Email: bkrishna@usc.edu.
  
    This work is sponsored in part by the Nature Science Foundation of China (No. 61701275, No. 91638204, No. 61571265, No. 61621091), the China Postdoctoral Science Foundation, and Hitachi Ltd. Part of the work has been submitted to IEEE International Symposium of Information Theory 2018 \cite{jiang18_isit}.
    }
    }
    \maketitle
    
    \begin{abstract}
    In a typical Internet of Things (IoT) application where a central controller collects status updates from multiple terminals, e.g., sensors and monitors, through a wireless multiaccess uplink, an important problem is how to attain timely status updates autonomously. In this paper, the timeliness of the status is measured by the recently proposed age-of-information (AoI) metric; both the theoretical and practical aspects of the problem are investigated: we aim to obtain a scheduling policy with minimum AoI and, meanwhile, still suitable for decentralized implementation on account of signalling exchange overhead. Towards this end, we first consider the set of arrival-independent and renewal (AIR) policies; the optimal policy thereof to minimize the time-average AoI is proved to be a round-robin policy with one-packet (latest packet only and others are dropped) buffers (RR-ONE). The optimality is established based on a generalized Poisson-arrival-see-time-average (PASTA) theorem. It is further proved that RR-ONE is asymptotically optimal among all policies in the massive IoT regime. The AoI steady-state stationary distribution under RR-ONE is also derived. A fully decentralized implementation of RR-ONE is proposed which can accommodate dynamic terminal appearances. In addition, considering scenarios where packets cannot be dropped, the optimal AIR policy thereof is also found with parameters given by the solution to a convex problem.
    \end{abstract}
    
    \section{Introduction}
    \label{sec_intro}
    Internet of Things (IoT) represents one of the biggest paradigm shifts recently which can revolutionize the information technology and several aspects of everyday-life such as living, e-health and driving; it envisions to transform every physical object into an intelligent individual that is capable of sensing, communicating and computing. By $2021$, Ericsson predicts that there will be around $28$ billion IoT devices and a large share of them are empowered by wireless communication technologies \cite{ericsson_iot}. 
    The wireless communication community has dedicated a significant amount of efforts to accommodate such a \emph{massive} number of emerging IoT devices; in particular, it is one of the main targets of the $5$G system that $100$-fold more connected devices per geographical area should be supported compared with current LTE systems \cite{oss14}. In addition to the sheer amount of IoT devices, it is also desired to enhance the timeliness of services for time-critical applications whereby the service quality depends heavily on the freshness of the monitoring data collected from IoT devices, e.g., the Tactile Internet and autonomous driving \cite{fet14}. 
    \begin{figure}[!t]
    \centering
    \includegraphics[width=0.5\textwidth]{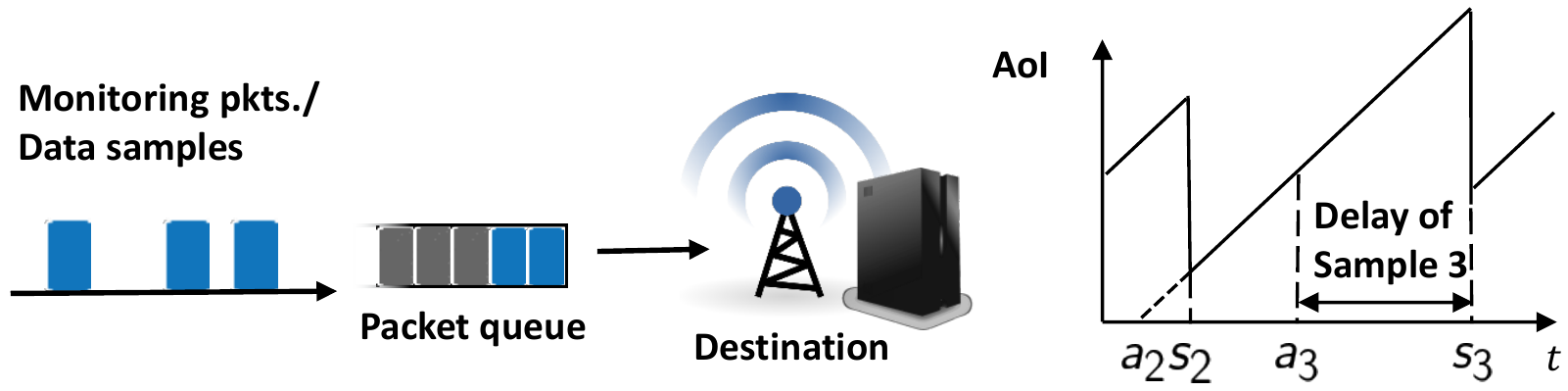}
    \caption{A status update example with one source (sampler), one destination and a single queue to illustrate AoI.}
    \label{fig_aoi}
    \end{figure}
    
    \emph{Age-of-information} (AoI) is a recently proposed metric specifically to quantify such timeliness \cite{kaul11,kaul12,yates12,yates16,kadota16,he17,joo17}; it is a constantly evolving information monitoring delay at a destination node, or simply put, time elapsed since the last-updated packet's generation. This definition jointly accounts for the delay introduced by sampling the information source and data communication, which distinguishes AoI from the conventional end-to-end communication (queuing and transmission) delay metric \cite{jiang17_iotj,neely10}. Consider a concrete example in Fig. \ref{fig_aoi}. The AoI only coincides with the communication delay at the time when a status update packet is delivered; another distinct difference is that the communication delay is defined for each packet; in contrast, the AoI is a constantly evolving measurement at the destination. The AoI characterizes the knowledge freshness of remote information sources at the destination. This knowledge and its timeliness are both essential for real-time applications. The wireless communication system is therefore well motivated to optimize the AoI, however, this objective may deviate from the conventional throughput- or delay-oriented paradigms since it has been shown that the optimization of AoI leads to distinctively different system designs, e.g., sampling strategy and service principle \cite{kaul12}. 

    One of the fundamental restrictions of wireless communication systems is that transmissions are subject to interference due to the broadcast nature of electromagnetic waves, leading to the fact that IoT terminals cannot transmit simultaneously; otherwise collisions happen and transmissions fail with no data delivered. Therefore, terminals should be carefully \emph{scheduled} to avoid such collisions; as a result, delay is introduced. For instance, a simple scheduling strategy is that terminals take turns to update their status data to avoid collisions. Intriguingly, we will show that taking turns (a round-robin scheduling policy with proper packet management) is, to some extent, the optimal policy without entailing a large amount of signalling overhead. 
    
    The overhead issue touches upon another important design principle in \emph{wireless multiaccess uplinks}, especially with massive distributed IoT devices, that is the policy design is preferably \emph{decentralized}, i.e., decisions are made autonomously at terminals and require only local information. For instance, the carrier-sensing-medium-access (CSMA) protocol is a widely-used and successful application of decentralized protocol in wireless networks \cite{bia00}. In particular, terminals transmit based on a contention protocol and scheduling decisions are made in a decentralized manner. However, the CSMA protocol is designed only for throughput maximization and may face severe challenges in status update systems. 
    
    Concerning the aforementioned scenario and corresponding challenges, the contributions of this paper include:
    
    \begin{itemize}
        \item 
        Among arrival-independent renewal (AIR) scheduling policies, whose decisions are independent with packet-arrival processes and hence decentralization-friendly, a round-robin policy with one-packet buffers (only retains the most up-to-date packet and others are discarded) at terminals (RR-ONE) is proved optimal. The proof technique leverages a generalized Poisson-arrival-see-time-average (PASTA) theorem which, as far as we know, has not been adopted in the related literature before. 
        \item
        RR-ONE is proved asymptotically optimal among \emph{all} policies with a massive number of terminals. It is shown that the optimum time-average AoI is proportional to the number of terminals asymptotically; the optimum linear scaling factor is $\frac{1}{2}$. RR-ONE is proved to achieve the optimum scaling factor. The CSMA protocol is however shown to have (at least) a scaling factor of $1$; hence its time-average AoI is arbitrarily larger than RR-ONE asymptotically. In addition, the AoI steady-state stationary distribution under RR-ONE is also derived.
        \item
        A full-fledged decentralized implementation of RR-ONE is described; it is capable of adapting to dynamic terminal appearances which is essential for decentralized algorithms. Thereby, the only global information required for each terminal is the total number of terminals; this is obtained by a common broadcast message from the central controller.
        \item
        Considering scenarios wherein arrival packets are queued and first-come-first-served (FCFS) at terminals without any packet management, e.g., packet dropping, the optimal AIR policy thereof is also found: it schedules every terminal based on deterministic intervals, with intervals given as the solution to a convex optimization problem; an approximate closed-form solution, which is shown to be optimal with heavy traffic, is also derived. 
    \end{itemize}
    
    The remainder of the paper is organized as follows. In Section \ref{sec_sm}, the system model is introduced; the problem of AoI minimization is then formulated; for clarity, we present our main results here and detailed proofs and explanations are conveyed in the subsequent sections. In Section \ref{sec_air}, we show that RR-ONE is the optimal AIR policy. In Section \ref{sec_rrone}, the asymptotic optimality is proved. In Section \ref{sec_sta}, the stationary distribution under RR-ONE is derived. In Section \ref{sec_dp}, a decentralized protocol is presented. Section \ref{sec_nopm} presents the optimal AIR policy without performing packet management. Section \ref{sec_sr} presents simulation results. Finally, in Section \ref{sec_cl}, conclusions are drawn and discussions are made. The proofs of several lemmas are shown in the appendix.
    \begin{figure}[!t]
    \centering
    \includegraphics[width=0.5\textwidth]{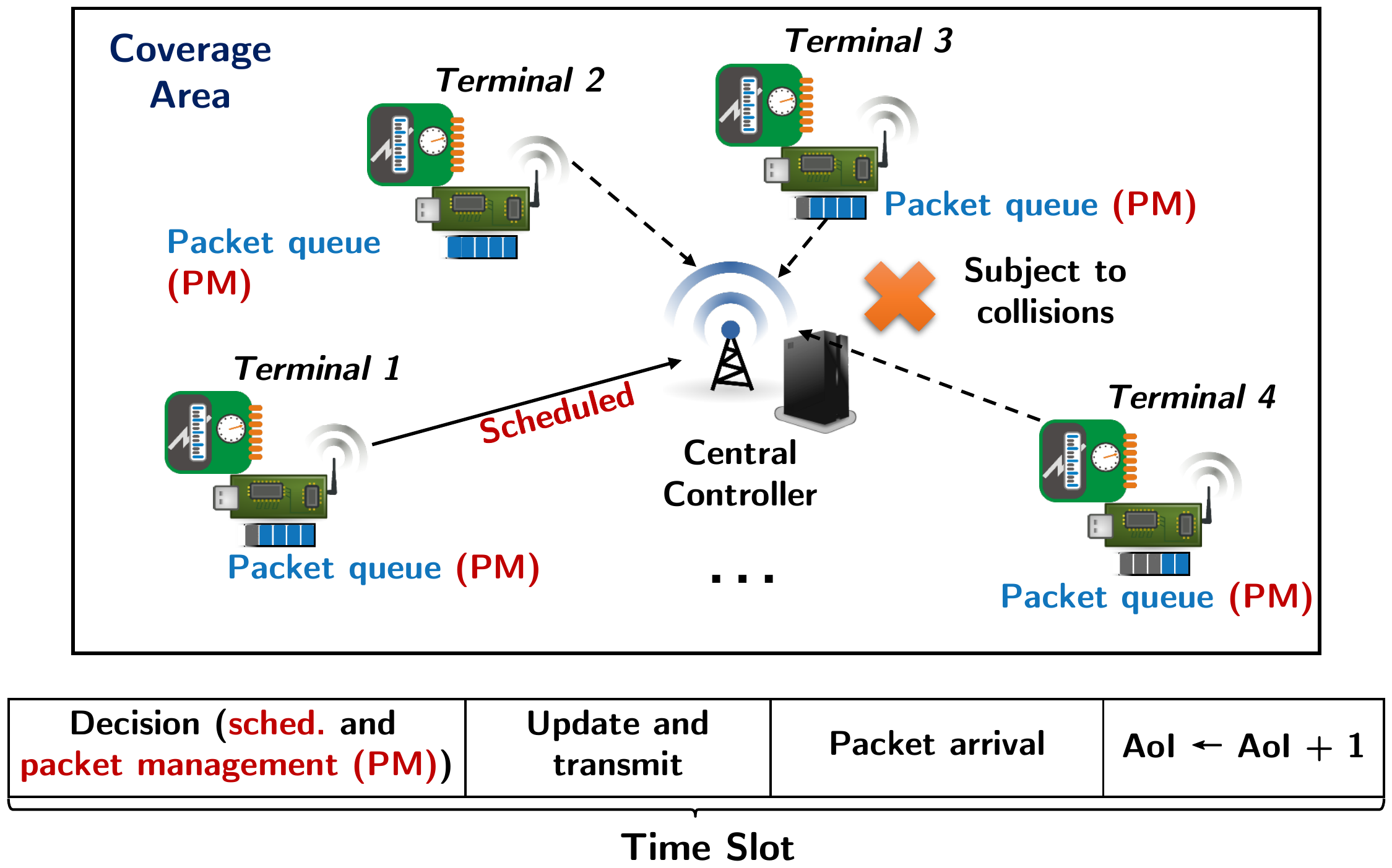}
    \caption{Considered system architecture and status update procedure. }
    \label{fig_arch}
    \end{figure}
    
    \subsection{Related Work}
    An AoI optimization problem can be posed as minimizing the time-average AoI at the receiver by controlling the sampling rate at the terminal. All the sampled data packets go through an $M/M/1$ queue as shown in Fig. \ref{fig_aoi}.\footnote{The AoI evolution in this paper is based on discrete time slots; whereas the AoI changes continuously in Fig. \ref{fig_aoi} and Fig. \ref{fig_asta} for ease of exposition.} There is a subtle, in most cases tradeoff, relationship between throughput and AoI: Sampling frequently results in a high throughput and a low sampling delay but may introduce a large queuing delay; on the other hand, sampling at a low rate introduces a large sampling delay, whereas the queuing delay is reduced. Queuing analysis reveals that there exists an optimal sampling rate, assuming the generated packets are queued and the service is FCFS \cite{kaul12}; $M/D/1$ and $D/M/1$ queues are also considered therein.
    
    Since its inception, AoI has received wide research attentions and there have been several extensions. Regarding multiple sources sharing one queue, and hence multiple corresponding AoIs, Yates and Kaul \cite{yates12,yates16} derive the minimum time-average AoI region under $M/M/1$ queuing model and different service disciplines, e.g., FCFS, last-come-fist-served (LCFS), LCFS with preemption. The service process can be extended to obey general distributions; in this regard, the average peak-AoI (PAoI) for multiple sources is obtained, and the problem of optimizing update rates is considered by Huang and Modiano \cite{huang15}. The scheduling problem for multiple servers and a single queue with job replication is considered by Bedewy \emph{et al.} \cite{bedewy16}. The AoI performance is shown to be improved by proper packet management, e.g., LCFS with preemption \cite{kaul12_ciss} and packet dropping mechanism \cite{costa16}, based on the intuition that the last-arrived packet has the least age. However, considering service interruption, always prioritizing a new packet may result in performance degradation \cite{najm16}. The optimization of sampling time without considering queuing delay is studied by Sun \emph{et al.} \cite{sun17}; they show that even without queuing delay the source should wait a certain time before sampling again to minimize time-average AoI due to service time uncertainty. 
    
    The study of the multi-queue scheduling problem is the most relevant work to ours \cite{kadota16,he17,hsu17}. Hsu \emph{et al.} approach this problem in the wireless broadcast channel where the scheduling decisions are centralized; they prove the optimal policy is age-threshold-based. The scheduling problem of multiple sources inside a finite-length transmission frame to minimize AoI is proved NP-hard \cite{he17}. Joo and Eryilmaz propose that the status updates from multiple sources should be synchronized in applications such as network monitoring and distributed sensing; they develop a drift-based approach to address the issue \cite{joo17}. The Whittle's index \cite{whitt84} is leveraged by Kadota \emph{et al.} \cite{kadota16} based on a restless multi-armed-bandit formulation; it is shown that an age-greedy policy is optimal in the symmetric case and the Whittle's index is derived for asymmetric cases. 
    
    The major distinction between our formulation and existing work is that we consider the following scenario: 1) status packets arrive at random time slots at terminals; 2) limited information is available for the policy decisions to facilitate decentralized implementation. The considered scenario is justified in massive IoT status update systems: status variation is unpredictable, and hence update packets, which are used to monitor status variation, are randomly generated; transmissions happen in a wireless multiaccess uplink where scheduling decisions are preferably decentralized to avoid overhead explosion. As far as we know, no existing work on AoI optimization has addressed a practical scenario involving both aspects.
    
    \section{System Model, Problem Formulation and Main Results}
    \label{sec_sm}
    Consider a base station (BS), alternatively referred to as central controller, which is responsible for collecting status update packets from a large number of IoT devices as shown in Fig. \ref{fig_arch}. A time-slotted system is considered. The status update packets are generated and stored at terminal queues. The queue buffer size for every terminal is assumed to be identical and denote by $B$. The number of packets generated at time $t$ of terminal $n$ is denoted by $L_n(t)$ and $L_n(t)$ is assumed to be a Bernoulli random variable with parameter $\lambda_n$; the arrival processes $\left\{L_n(t),\,t=1,2,... \right\}$, $n \in \{1,...,N\}$ are independent over terminals and time. The number of terminals is denoted by $N$. Let $U_{n,\pi}(t)$ denote the scheduling decision of terminal $n$ at time $t$ for a given policy $\pi$, i.e., $U_{n,\pi}(t)=1$ if terminal $n$ is scheduled and $U_{n,\pi}(t)=0$ otherwise.
    
    The $T$-horizon time-average AoI is denoted by
    \begin{equation}
    \label{AoI}
        \bar{h}_{\pi}^{(T,N)} \triangleq \frac{1}{TN}\sum_{t=1}^T \sum_{n=1}^N h_{n,\pi}(t),
    \end{equation}
    where the AoI at the $t$-th time slot for terminal $n$ based on policy $\pi$ is denoted by $h_{n,\pi}(t)$, and the time horizon is $T$ time slots. Denote time-average AoI over infinite time horizon as 
    \begin{equation}
    \label{aoi_inf}
        \bar{h}_{\pi}^{(\infty,N)} \triangleq \lim_{T \to \infty} \bar{h}_{\pi}^{(T,N)}. 
    \end{equation}
    The evolution of AoI can be written as 
    \begin{equation}
    \label{evo}
        h_{n,\pi}(t+1) = h_{n,\pi}(t) - U_{n,\pi}(t) \prod_{m \neq n}(1-U_{m,\pi}(t))g_n(t) + 1,
    \end{equation}
    where $g_n(t)$ denotes the AoI reduction with a successful update from terminal $n$. Consequently, we have $g_n(t)=0$ when queue-$n$ is empty at time $t$. The AoI for each terminal always increases by one after each time slot. Based on this definition \eqref{evo}, whenever a collision happens, i.e., more than one terminals transmit in the same time slot, no status is updated. Note that transmission failures only happen with collisions, otherwise the transmission is always assumed successful; this corresponds to the interference-limited regime which is emerging to be the main application scenario in the future ultra-dense networks \cite{Andrews14}, and therein failures due to noise is negligible. In addition, denote the time-average AoI of terminal-$n$ under policy $\pi$ as 
    \begin{equation}
        \bar{h}_{n,\pi}^{(T)}  \triangleq \frac{1}{T}\sum_{t=1}^T h_{n,\pi}(t).
    \end{equation}
    
    The status update procedure is described in Fig. \ref{fig_arch}. We assume the following sequence of events in each time slot. At the beginning of each time slot, scheduling decisions are made, including:
    \begin{itemize}
        \item 
        \emph{Terminal scheduling}: Decide which terminal, or a set of terminals\footnote{This case is described only for completeness and not considered in this paper since it results in a collision.}, updates and transmits in this time slot. 
        \item
        \emph{Packet management}: Once scheduled, the terminal can apply a packet management scheme, e.g., it can choose a packet from its queue to update in the time slot, or drop several packets. In contrast, packets are queued and served based on an FCFS manner without any packet management.  
    \end{itemize}
    Based on the decision, the scheduled terminal transmits its update packet (assuming one packet is transmitted in each time slot), and thereby the AoI is refreshed at the BS. Afterwards, packets arrive randomly at terminals (the age of newly arrived packets is zero) and then the ages of all packets and all AoIs increase by one. This marks the end of a time slot in our model. The AoI at one time slot is defined as the AoI at the end of the time slot after status update and natural growth. 
    
    The objective considered in this paper is to minimize the infinite-horizon time-average AoI \eqref{aoi_inf} over all policies.\footnote{Hereinafter, we refer to \eqref{aoi_inf} as time-average AoI for simplicity.} As a first step, the following definition and Lemma \ref{lm_0} (cf. proof in Appendix \ref{app_lemmas}) enable us to only consider work-conserving non-collision (WCNC) policies without loss of optimality. 
    
    \begin{defi}[WCNC policy]
    A WCNC policy is defined as a policy that is not idle when there is at least one packet in terminal queues, nor schedule more than one terminals simultaneously. $\hfill\square$
    \end{defi}    
    \begin{lemma}
    \label{lm_0}
    For a non-WCNC policy, there exists at least one WCNC policy that achieves lower AoI. $\hfill\square$
    \end{lemma}
    
    For practical concerns that the policy decisions should be decentralized, and also mathematical tractability, consider AIR policies defined as follows. Denote the resultant scheduling interval process of terminal-$n$ based on policy $\pi$ as $X_{n,\pi}^{(k)}$, $k=1,2,...$ where $k$ is the scheduling interval index. Define $R_{n,\pi}(t)$ as the counting process of scheduling times before time $t$ for terminal $n$, i.e., $R_{n,\pi}(t) \triangleq \textrm{sup} \{r: \sum_{k=0}^r X_{n,\pi}^{(k)} \le t\}$.
    \begin{defi}[AIR policy]
    \label{def_air}
    A policy $\pi$ is an AIR policy if the following conditions are both met.
    \begin{enumerate}[label=(\roman*)]
        \item
        \label{cond1}
        The scheduling interval processes $\{X_{n,\pi}^{(k)},\,n=1,...,N\}$ are independent with the packet arrival processes at terminals, with finite first and second raw moments denoted by $m_n$ and $v_n$ respectively.
        \item
        \label{cond2}
        The counting processes $R_{n,\pi}(t)$, $n=1,...,N$ are renewal processes.  $\hfill\square$
    \end{enumerate}   
    \end{defi}
    
    By definition, the set of AIR policies is essentially a subset of all policies. The condition (i) is in fact reflecting the practical perspective that the scheduling decisions are desired to be independent of the packet arrival processes to enable decentralized implementation and reduce signalling exchange overhead. The condition (ii) does enforce an additional constraint that the scheduling intervals are i.i.d.; however the distributions can be arbitrary as long as they have finite first and second moments. Note that, notwithstanding these conditions, it is found (Theorem \ref{thm1}) that the optimal AIR policy with proper packet management is asymptotically optimal among all policies in the massive IoT regime. 
    
    \subsection{Main Results}        
    \begin{figure}[!t]
    \centering
    \includegraphics[width=0.45\textwidth]{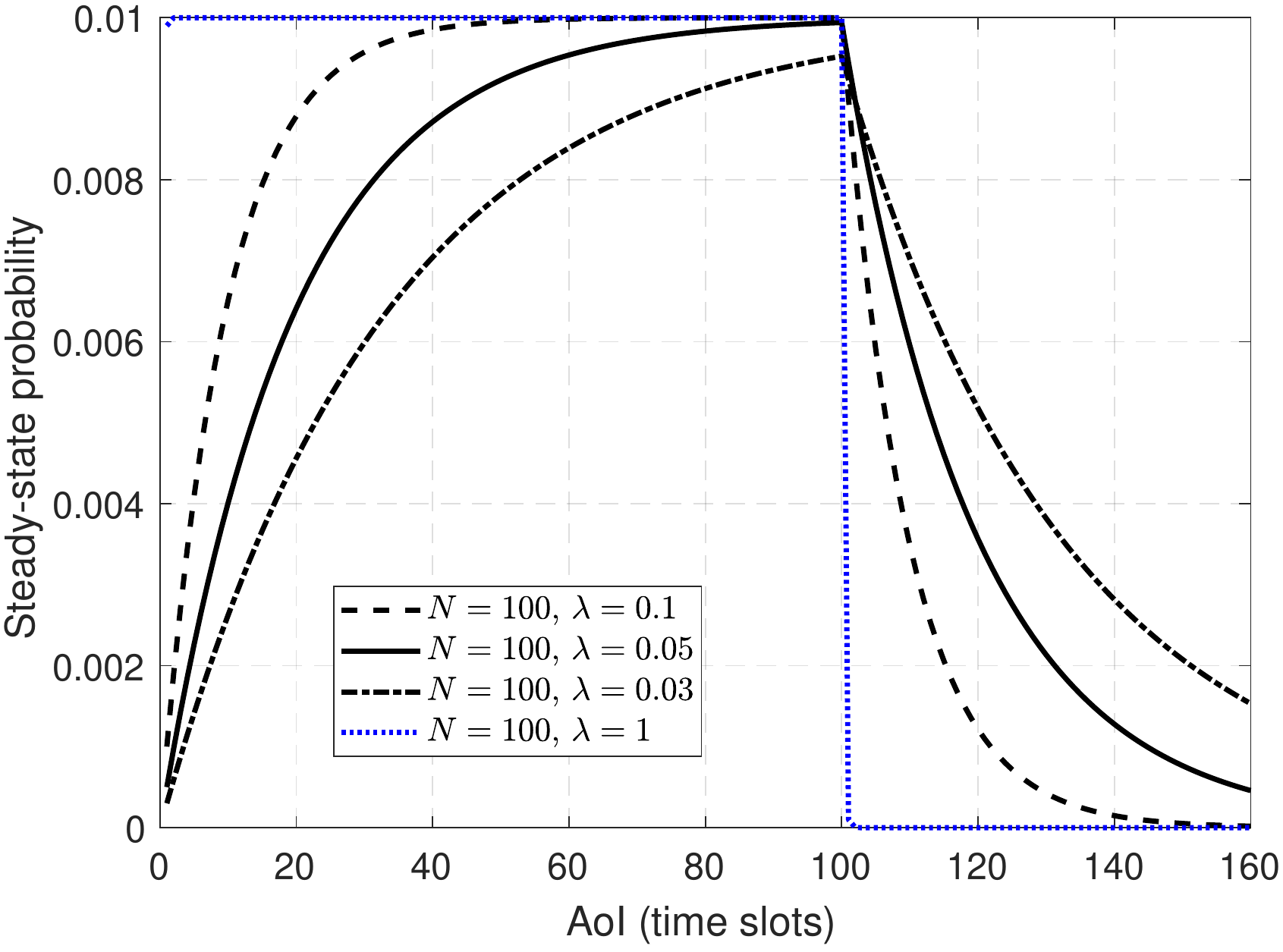}
    \caption{Steady-state stationary distribution of AoI under RR-ONE. The blue dotted curve represents a uniform distribution on $\{1,...,N\}$ when $\lambda=1$ since in this case the AoI evolves periodically from $1$ to $N$ (Lemma \ref{lma0}).}
    \label{fig_dist}
    \end{figure}
    \begin{defi}[RR-ONE]
    RR-ONE, denoted by $\mathsf{RR}$ in the subscript, is defined as a policy that schedules the $n_\mathsf{RR}$-th terminal at each time slot which satisfies
    \begin{equation}
        n_\mathsf{RR} = \min\left\{n: \tau_n = \max_{m=1,...,N} \tau_m\right\},
    \end{equation}
    and only retains the last-arrival packet at each terminal. The time since last update from terminal $m$ is denoted by $\tau_m$. $\hfill\square$
    \end{defi}
    
    \begin{theorem}
    \label{thm_air}
    RR-ONE is the optimal AIR policy to minimize the time-average AoI, with
    \begin{IEEEeqnarray}{rCl}
        \bar{h}_{\mathsf{RR}}^{(\infty,N)} = \frac{1}{N} \sum_{n=1}^N \frac{1}{\lambda_n} + \frac{N-1}{2}.
    \end{IEEEeqnarray}
    $\hfill\square$
    \end{theorem}
    
    \begin{theorem}
    \label{thm1}
    RR-ONE is asymptotically optimal among all policies in the massive IoT regime; it achieves the optimum asymptotic scaling factors, i.e.,
    \begin{equation}
    \label{nlim}
        \lim_{N \to \infty} \frac{\bar{h}_{\mathsf{RR}}^{(\infty,N)}}{N} = \lim_{N \to \infty} \frac{\bar{h}_{\textrm{opt}}^{(\infty,N)}}{N} = \frac{1}{2},\,\forall \lambda_i,
    \end{equation}
    and
    \begin{IEEEeqnarray}{rCl}
    \label{llim}
        \lim_{\frac{1}{\lambda_i} \to \infty} \lambda_i \bar{h}_{\mathsf{RR}}^{(\infty,N)} &=& \lim_{\frac{1}{\lambda_i} \to \infty} \lambda_i \bar{h}_{\textrm{opt}}^{(\infty,N)} =  \frac{1}{N},\nonumber\\
        && \forall \lambda_{j,j\neq i}\textrm{ and }N,
    \end{IEEEeqnarray}
    where $\bar{h}_{\textrm{opt}}^{(\infty,N)}$ denotes the minimum time-average AoI. $\hfill\square$
    \end{theorem}
    
    \begin{coro}
    The minimum time-average AoI is proportional to the number of terminals asymptotically, and the optimum scaling factor is $\frac{1}{2}$. $\hfill\square$
    \end{coro}
    
    We emphasize that the asymptotic optimality of RR-ONE is among all policies, including policies requiring global information, e.g., terminals' queue lengths, age of all packets, and even non-causal policies which know future packet arrivals (by observing that the proof of Theorem \ref{thm1} also applies in this case).
    
    \begin{theorem}
    \label{thm3}
    The AoI evolution of terminal-$n$ based on RR-ONE follows a Markov renewal process with a fixed renewal time of $N$ time slots, and the steady-state stationary distribution is
    \begin{equation}
    \label{eq_dist}
        \mu_{n}(j) = \left\{\,
        \begin{IEEEeqnarraybox}[][c]{l?s}
        \IEEEstrut	
    	\frac{1-(1-\lambda_n)^j}{N}, & $1 \le j \le N$;\\
    	\frac{(1-\lambda_n)^{j-N}}{N}(1-(1-\lambda_n)^N), & $j \ge N+1$, 
    	\IEEEstrut
    	\end{IEEEeqnarraybox}
    	\right.
    \end{equation}
    where $\mu_{n}(j)$ denotes the probability that the steady state AoI of terminal-$n$ is $j$.  $\hfill\square$
    \end{theorem}
    
    The steady-state stationary distribution in \eqref{eq_dist} is instantiated, and an insight is described in Fig. \ref{fig_dist}.     
    
    \begin{theorem}
    \label{thm2}
    The optimal AIR policy \emph{without packet management} to minimize the time-average PAoI \eqref{paoi} is to schedule each terminal based on a deterministic scheduling interval\footnote{This may not be possible given the fact that the optimal scheduling intervals can be fractional numbers, and that different terminals' scheduling arrangements may conflict with each other. Nevertheless, this complication can be addressed by adopting a policy that approximately satisfies the conditions with minimal performance degradation.}. The scheduling interval for terminal-$n$ is given by $m^*_n = \frac{1}{\beta^*_n}$ where $\beta^*_n$ is the solution of the convex optimization problem:
    \begin{flalign}
    \label{p2}
    \textbf{P2:}&&\mathop{\textrm{minimize}}\limits_{\beta_n\,n=1,...,N}  \,\,& \frac{1}{N} \sum_{n=1}^N \left[\frac{1}{\lambda_n} + \frac{1}{2}\left(\frac{1}{\beta_n-\lambda_n}+\frac{1}{\beta_n}\right)\right]  &&\nonumber\\
    &&\textrm{s.t.,}\,\, &  \beta_n > \lambda_n,\,\forall n=1,...,N,  &&\nonumber\\
    &&& \sum_{n=1}^N \beta_n = 1. &&
    \end{flalign}
    The corresponding policy is referred to as RR-$\bm{\lambda}$. $\hfill\square$
    \end{theorem}
    
    An important note is that both RR-ONE and RR-$\bm{\lambda}$ can be easily adapted for decentralized implementation. Roughly illustrated for RR-ONE, each terminal is assigned a unique time slot to transmit in a frame of length $N$, and only retains the most up-to-date packet in its buffer. A detailed protocol which accounts for variable $N$, i.e., random terminal appearances, is described in Section \ref{sec_dp}. Although RR-$\bm{\lambda}$ requires the statistical knowledge of packet arrival rates of terminals, it is independent with the instantaneous system state and therefore the overhead is acceptable.
    In the following sections, we will prove our main results and elaborate on their implementations and implications.
    \section{Proof of Theorem \ref{thm_air}: Optimal AIR Policy}
    \label{sec_air}
    The quest for the optimal policy, among all policies with any given $N$, to minimize the time-average AoI seems elusive, because the problem can be essentially viewed as a restless multi-armed bandit problem with time- and arm-correlated reward functions \cite{ahm09}. Besides, there is a strong probability that the optimal policy requires global information exchange and hence decentralization-unfriendly. Therefore, in this section, we resolve to derive the optimal AIR policy to minimize the time-average AoI in \eqref{aoi_inf} following a generalized Poisson-arrival-see-time-average (PASTA) theorem, i.e., the arrival-see-time-average (ASTA) property with a Markov state process as the observed process and an independent outside observer \cite{ben95}. First, consider the queue evolution of terminal-$n$ based on AIR policies; it is similar with an $M/G/1$ queue given the definition of AIR policies, with a subtle, but important, difference that the service (in this case the service time is the scheduling interval) begins immediately after a packet departure, even if there is no packet waiting in the queue. In the case that there is no packet is in the queue, the service proceeds independently till the end, i.e., scheduled, during which period two possible circumstances can occur: 1) there are (at least one) packet arrivals and thereby one of the packets is updated under a certain packet management policy; 2) there is no packet arrival and consequently no packet is updated. It is clear that under this queue model, the optimal packet management, under arbitrary scheduling policy, is to always update the most up-to-date packet, i.e., the packet that arrives the last; the resultant queue is equivalent to having a buffer size of one and storing only the latest arrival packet. Note that this packet management policy is not necessarily optimal with preemptive service model due to service interruption \cite{najm17}. Without loss of optimality, we only consider the one-packet buffer packet management policy in the rest of the section.
    
\begin{figure*}[!t]
    \centering
    \includegraphics[width=0.65\textwidth]{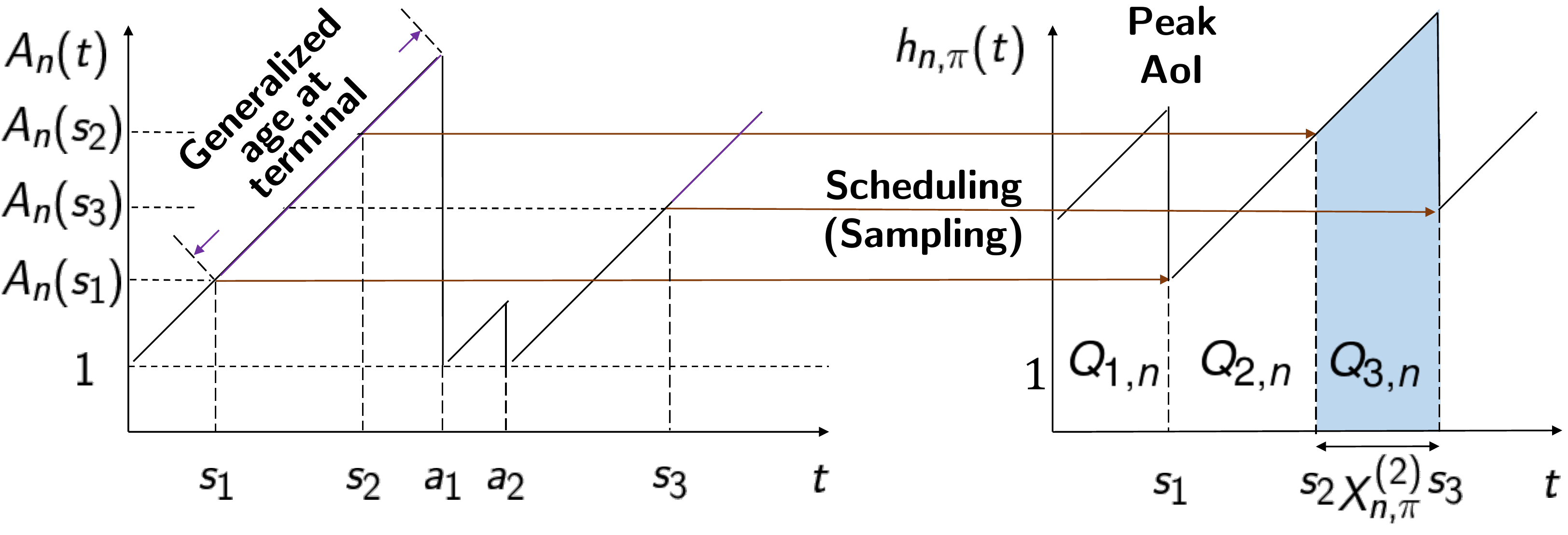}
    \caption{Age of the packet at terminal-$n$ assuming one packet buffer (left) and AoI at the BS (right).}
    \label{fig_asta}
    \end{figure*}

    The age of the packet in queue-$n$ (buffer size is one) is denoted by $A_n(t)$, $t=1,2,...$, a sample path of which is shown in the left of Fig. \ref{fig_asta}. Upon a packet arrival, e.g., $a_i$ in Fig. \ref{fig_asta}, the age $A_n(t)$ drops to one (measured at the end of the time slot) based on the procedure in Fig. \ref{fig_arch}. When terminal-$n$ is scheduled at the time of $s_i$, the AoI at the BS is updated to the age of the packet at terminal $n$, i.e., $A_n(s_i)$. Note that we prescribe a generalized age of $A_n(s_i)$ that between each update and next packet arrival, e.g., between $s_1$ and $a_1$, $A_n(t)$ equals the AoI of terminal $n$ at the BS although there is no packet in the queue during the time. By doing this, we make $A_n(t)$ evolve \emph{independently} with $h_{n,\pi}(t)$ while not affecting the AoI update procedure; this is crucial for the ASTA property to apply. 
    
    Based on the renewal process condition of AIR policies, and following the same arguments in, e.g., \cite{najm17}, the time-average AoI can be readily calculated by the sum of the geometric areas $Q_{k,n}$ in Fig. \ref{fig_asta}:
    \begin{IEEEeqnarray}{rCl}
    \label{aoi_tr}
        \bar{h}_{n,\pi}^{(\infty,N)} = \lim_{T \to \infty} \frac{K}{T} \frac{1}{K} \sum_{k=1}^{K} Q_{k,n} &=& \lim_{T \to \infty} \frac{K}{T} \lim_{K \to \infty} \frac{1}{K} \sum_{k=1}^{K} Q_{k,n} \nonumber\\
        &=& \frac{\mathbb{E}[Q_{k,n}]}{m_n}. 
    \end{IEEEeqnarray}
    The last equality is based on the elementary renewal theorem \cite{cox67}. It then follows that
    \begin{IEEEeqnarray}{rCl}
    \label{cons_cond}
        \bar{h}_{n,\pi}^{(\infty,N)} &=& \frac{1}{m_n} \mathbb{E}\left[X_{n,\pi}^{(k)} A_n(s_k) + \left(X_{n,\pi}^{(k)}-1\right)\frac{X_{n,\pi}^{(k)}}{2}\right] \nonumber\\
        &\overset{(a)}{=}& \frac{1}{m_n} \left(\mathbb{E}\left[X_{n,\pi}^{(k)}\right] \mathbb{E}\left[A_n(s_k)\right] + \frac{1}{2} (v_n-m_n)\right) \nonumber\\
        & = & \mathbb{E}\left[A_n(s_k)\right] + \frac{v_n-m_n}{2m_n} \nonumber\\
        &\overset{(b)}{\ge}& \mathbb{E}\left[A_n(s_k)\right] + \frac{m_n-1}{2},
    \end{IEEEeqnarray}
    where the equality $(a)$ is based on the arrival-independent condition of AIR policies, and the inequality $(b)$ follows from $v_n \ge m_n^2$; note that the equality holds when the scheduling interval is a constant $m_n$. 
    
    It is now clear that the main challenge is to calculate $\mathbb{E}\left[A_n(s_k)\right]$. First we have Lemma \ref{lm_1} (cf. proof in Appendix \ref{app_lemmas}) which shows that $\left\{A_n(t),\,t=1,2...\right\}$ is a Markov state process and its stationary distribution is given in \eqref{46}.    
    \begin{lemma}
    \label{lm_1}
    $\left\{A_n(t),\,t=1,2...\right\}$ is a Markov state process with the steady-state stationary distribution given as
    \begin{equation}
    \label{46}
        \mu_n(j) = \lambda_n (1-\lambda_n)^{j-1},
    \end{equation}
    where $\mu_n(j)$ denotes the probability that the steady state of terminal $n$ is state $j$ (age of packet at terminal-$n$ equals $j$). $\hfill\square$
    \end{lemma}   
    
    Then the challenge of calculating $\mathbb{E}\left[A_n(s_k)\right]$ is tackled by treating $\mathbb{E}\left[A_n(s_k)\right]$ as the average state value of a Markov state process by an independent outside observer. Armed with this, we invoke the ASTA property \cite{ben95} which can be seen as a generalization of the well-known PASTA theorem to non-Poisson observers.
    \begin{lemma}
    \label{lm_asta}
    \cite[Theorem 3.14]{ben95} Let $U$ be a Markov state process and $N$ be a counting process. Then ASTA holds for the pair $(U,N)$ if $U$ is left-continuous and the pair $(U,N)$ is forward-pointwise independent, i.e., for all $t>0$, $U(t)$ and $\{N(t+s)-N(s):\,s \ge 1\}$ are independent. $\hfill\square$
    \end{lemma}
    
    Let $U$ be $\left\{A_n(t),\,t=1,2...\right\}$, and $N$ in Lemma \ref{lm_asta} be the counting process of the number of scheduling times before time $t$. Then based on the AIR policy conditions, $U$ and $N$ are independent. The continuous condition follows by design of update sequence described in Fig. \ref{fig_arch}. Therefore, we obtain
    \begin{equation}
        \mathbb{E}\left[A_n(s_k)\right] = \lim_{K \to \infty} \frac{1}{K} \sum_{k=1}^{K} A_n(s_k) = \mathbb{E}\left[A_n(t)\right] = \frac{1}{\lambda_n}.
    \end{equation}
    In other words, the time-average of random sampling ($s_k$) of the Markov process $\left\{A_n(t),\,t=1,2...\right\}$ equals the steady-state average. Combining with \eqref{cons_cond}, the time-average AoI is 
    \begin{IEEEeqnarray}{rCl}
    \label{rate_cond}
        \bar{h}_{\pi}^{(\infty,N)} = \frac{1}{N} \sum_{n=1}^N \bar{h}_{n,\pi}^{(\infty,N)} &\ge&  \frac{1}{N} \sum_{n=1}^N \left(\mathbb{E}\left[A_n(s_k)\right] + \frac{m_n-1}{2}\right) \nonumber\\
        &=&  \frac{1}{N} \sum_{n=1}^N \left(\frac{1}{\lambda_n} + \frac{m_n-1}{2}\right).
    \end{IEEEeqnarray}
    The update rate of all terminals equals one packet per time slot for WCNC policies; therefore, according to the elementary renewal theorem, 
    \begin{equation}
        \sum_{n=1}^N \frac{1}{m_n}=1.
    \end{equation}
    It follows from the arithmetic-harmonic-mean inequality (equality holds when $m_n=N$, $\forall n = 1,...,N$) and \eqref{rate_cond} that
    \begin{equation}
    \label{eq_1}
        \bar{h}_{\pi}^{(\infty,N)} \ge \frac{1}{N} \sum_{n=1}^N \frac{1}{\lambda_n} + \frac{N-1}{2}.
    \end{equation}
    
    The equality holds in \eqref{eq_1} under two conditions: 1) $m_n=N$, $\forall n = 1,...,N$; 2) the scheduling interval is a constant $m_n$. These two conditions can be both satisfied with RR-ONE. For a sanity check, RR-ONE is indeed an AIR policy. With this, we can conclude the proof of Theorem \ref{thm_air}.
    \begin{remark}
    Alert readers may be curious about the optimal policy in general. In particular, since a myopic policy (minimize the AoI in the next time slot) is optimal \cite[Theorem 1]{kadota16} when we eliminate the randomness of arrival packets ($\lambda=1$), i.e., the AoI is updated to one (or zero) every time, is it optimal in general? A counter example is given in Appendix \ref{app1} to show that the myopic policy is not optimal even with global state information (GSI), e.g., queue length, age of packets and arrival rates information, at least with finite horizon. Nevertheless, we conjecture that the myopic policy with GSI is close to optimal with infinite horizon, and it is adopted as a performance benchmark in Section \ref{sec_sr} to investigate the value of GSI.   
    \end{remark}

    \section{Proof of Theorem \ref{thm1}: Asymptotic Optimality}
    \label{sec_rrone}
    In IoT systems, one major challenge is to accommodate a large number of terminals while maintaining timely status updates. Hence, it is of central interest to consider the problem in the asymptotic regime. Towards this end, it will be shown in the following that RR-ONE, given its simple structure, is asymptotically optimal among all policies with arbitrary information and even non-causal packet arrival knowledge. First, we obtain two performance bounds and compare these with the achievable performance by RR-ONE; the conclusion follows by showing that they have identical asymptotic scaling factors.
 
    First, we introduce two lower bounds of the time-average AoI of any policies in Lemma \ref{lma0} and Lemma \ref{lma1} (cf. proof in Appendix \ref{app_lemmas}). 
    \begin{lemma}
    \label{lma0}
    The time-average AoI in \eqref{aoi_inf} cannot be less than $\frac{N+1}{2}$, i.e.,
    \begin{equation}
        \bar{h}_{\pi}^{(\infty,N)} \ge \frac{N+1}{2},\, \forall N=1,2,...,\, \lambda_n \in [0,1], n \in \{1,...,N\}.
    \end{equation}
    $\hfill\square$
    \end{lemma}
    
    \begin{lemma}
    \label{lma1}
    The time-average AoI in \eqref{AoI} cannot be less than $\frac{1}{N} \sum_{n=1}^N \frac{1}{\lambda_n}$, i.e.,
    \begin{IEEEeqnarray}{rCl}
        \bar{h}_{\pi}^{(\infty,N)} \ge \frac{1}{N} \sum_{n=1}^N \frac{1}{\lambda_n},\, &\forall& N=1,2,...,\, \lambda_n \in [0,1],\nonumber\\
        && n \in \{1,...,N\}.
    \end{IEEEeqnarray}
    $\hfill\square$
    \end{lemma}
    
    It follows that the minimum time-average AoI, denoted by $\bar{h}_{\textrm{opt}}^{(\infty,N)}$, cannot be less than either bound, i.e.,
    \begin{equation}
        \bar{h}_{\textrm{opt}}^{(\infty,N)} \ge \max\left[\frac{N+1}{2},\,\frac{1}{N}\sum_{n=1}^N \frac{1}{\lambda_n}\right].
    \end{equation}
    
    After obtaining two lower bounds in Lemma \ref{lma0} and \ref{lma1}, combining with the achievable AoI derived in \eqref{eq_1}, we can prove the asymptotic optimality of RR-ONE; the optimum scaling result follows immediately. Based on Lemma \ref{lma0}, Lemma \ref{lma1} and Theorem \ref{thm1}, it follows that $\forall N, \lambda_1,...,\lambda_N$,
    \begin{equation}
    \label{lim}
        \max\left[\frac{N+1}{2},\,\frac{1}{N}\sum_{n=1}^N \frac{1}{\lambda_n}\right] \le \bar{h}_{\textrm{opt}}^{(\infty,N)}  \le \frac{1}{N} \sum_{n=1}^N \frac{1}{\lambda_n} + \frac{N-1}{2}.
    \end{equation}
    For any fixed $\lambda_n$, $n=1,...,N$, divide both sides of \eqref{lim} by $N$, and let $N$ goes to infinity, we obtain
    \begin{IEEEeqnarray}{rCl}
        &&\lim_{N \to \infty} \frac{\max\left[\frac{N+1}{2},\,\frac{1}{N}\sum_{n=1}^N \frac{1}{\lambda_n}\right]}{N} =\frac{1}{2}, \nonumber\\
        &&\lim_{N \to \infty} \frac{\frac{1}{N} \sum_{n=1}^N \frac{1}{\lambda_n} + \frac{N-1}{2}}{N}=\frac{1}{2},
    \end{IEEEeqnarray}
    and therefore \eqref{nlim} follows. The scaling results for RR-ONE follows directly from \eqref{rr_aoi}. Similarly $\forall n$,
    \begin{IEEEeqnarray}{rCl}
        &&\lim_{\frac{1}{\lambda_n} \to \infty} \frac{\max\left[\frac{N+1}{2},\,\frac{1}{N}\sum_{n=1}^N \frac{1}{\lambda_n}\right]}{\frac{1}{\lambda_n}} =\frac{1}{N}, \nonumber\\
        &&\lim_{\frac{1}{\lambda_n} \to \infty} \frac{\frac{1}{N} \sum_{n=1}^N \frac{1}{\lambda_n} + \frac{N-1}{2}}{\frac{1}{\lambda_n}}=\frac{1}{N},
    \end{IEEEeqnarray}
    and \eqref{llim} follows, which concludes the proof of Theorem \ref{thm1}.
    \begin{remark}
    Theorem \ref{thm1} shows that the optimum time-average AoI scales linearly with the number of terminals $N$ and expected inter-arrival time, i.e., $\frac{1}{\lambda_n}$, $\forall n$. The optimum scaling factors are also given. Moreover, we show that RR-ONE not only can achieve linear scaling, but also achieves the optimum scaling factors. 
    \end{remark}
    \begin{remark}
    \label{one_remark}
    By setting $\lambda_n=1$, $\forall n$ in Theorem \ref{thm1} which is equivalent to a scenario wherein the age after every update is always one, it is immediately obvious that RR-ONE is optimal in this setting with arbitrary $N$.
    \end{remark}
    \begin{coro}
    \label{coro_un}
    The time-average AoI achieved by a uniformly random scheduling policy with one-packet buffers (UN-ONE) is at least
    \begin{equation}
    \label{un}
       \bar{h}_{\mathsf{UN}}^{(\infty,N)} \ge N. 
    \end{equation}
    \end{coro}
    \begin{IEEEproof}
    Consider running the UN-ONE policy in system $\mathsf{A0}$ (Lemma \ref{lma0}). Obviously this gives us a lower bound on the UN-ONE performance. In this case, the system state is fully characterized by the AoI of each terminal, and the AoI for each terminal (omitting the terminal index for brevity) evolves as 
    \begin{equation}
        p_{h,h+1} = \left(1-\frac{1}{N}\right),\,p_{h,1} = \frac{1}{N},\,p_{h,i}=0,\forall i \neq 1, h+1.
    \end{equation}
    Note that although the AoI transitions of different terminals are not independent by observing, e.g., only one terminal can be scheduled at each time slot, the time-average AoI in \eqref{aoi_inf} only concerns with the marginal distribution of AoI for each terminal. The AoI transition Markov chain of one terminal is shown in Fig. \ref{fig_un}. 
    \begin{figure}[!h]
    \centering
    \includegraphics[width=0.45\textwidth]{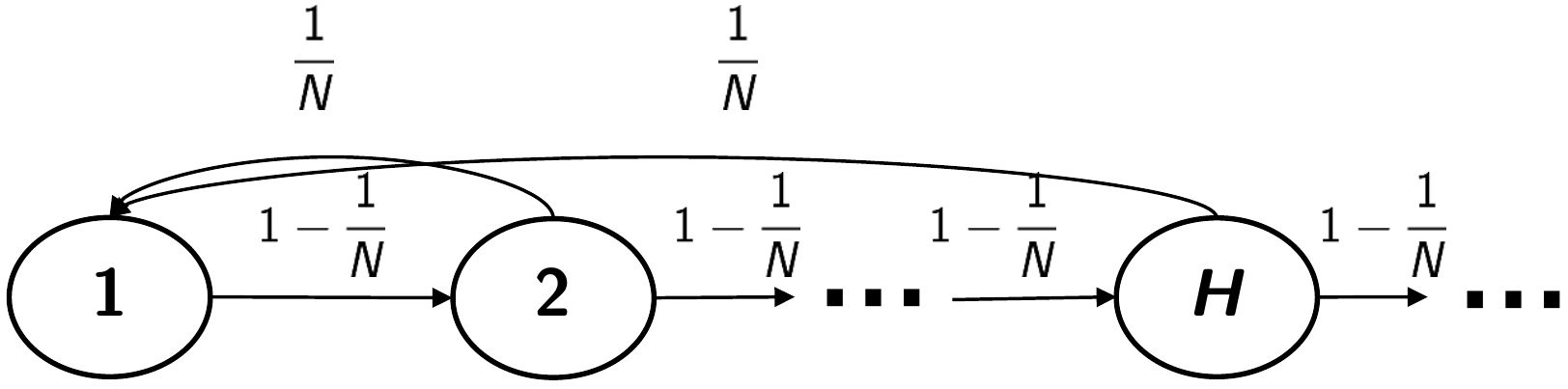}
    \caption{AoI transition Markov chain under UN-ONE in $\mathsf{A0}$.}
    \label{fig_un}
    \end{figure}
    
    Therefore, the steady state AoI for each terminal is exponentially distributed with parameter $\frac{1}{N}$. The time-average AoI by running the UN-ONE policy in system $\mathsf{A0}$ is hence $N$. Therefore, \eqref{un} follows immediately.  
    \end{IEEEproof}
    \begin{remark}
    Based on Corollary \ref{coro_un}, UN-ONE, which in fact can be seen as a performance lower bound of CSMA scheme without considering the contention time overhead, has a much larger time-average AoI compared with RR-ONE. In particular, when the number of terminals grows large, UN-ONE does not achieve the optimum scaling factor and thus is arbitrarily worse than RR-ONE.
    \end{remark}

    \section{Proof of Theorem \ref{thm3}: AoI Stationary Distribution under RR-ONE}
    \label{sec_sta}
    Although the time-average AoI under RR-ONE has been derived in Section \ref{sec_air} based on the ASTA property, the steady-state stationary distribution of AoI under RR-ONE is still unclear. To address this issue, it is found that the AoI evolution for each terminal follows a Markov renewal process with a fixed renewal interval of $N$ time slots. Therefore, the steady-state stationary distribution can be derived by first studying the embedded Markov chain. Denote the time since last update of the $n$-th terminal as $\tau_n$, it follows that the age of the packet at the $n$-th terminal, i.e., $A_n$ (the time index is omitted for brevity), conditioned on $\tau_n$ is distributed as (let $A_n = h_n$ denote there is no packet in the $n$-th queue at the time and thus the age after this update does not change)
    \begin{equation}
    \label{a_tau}
        \Pr \left\{A_n = a | \tau_n = \tau \right\} = \lambda_n(1-\lambda_n)^{a}, \, a={0,...,\tau-1}, \nonumber
    \end{equation}
    \begin{equation}
    \label{yn}
        \Pr\left\{A_n = h_n | \tau_n = \tau \right\} = (1-\lambda_n)^{\tau}. 
    \end{equation}
    If the terminal-$n$ is scheduled in this time slot, then the AoI of terminal-$n$, i.e., $h_n$, is updated to the realization of $A_n$. 
    
    Consider $N$ embedded Markov chains which describe the AoI transition for $N$ respective terminals between successive scheduling based on RR-ONE. The state $S$ of the $n$-th Markov chain (the terminal index is omitted for brevity) is defined as the AoI at the scheduled time slot. The transition probability is therefore  
    \begin{IEEEeqnarray}{rCl}
    && p_{ss^\prime} \triangleq \Pr\left\{S_{k+1} = s^\prime|S_k=s\right\} \nonumber\\
    &=& \left\{\,
        \begin{IEEEeqnarraybox}[][c]{l?s}
        \IEEEstrut	
    	\lambda_n (1-\lambda_n)^{s^\prime-1}, & if $s^\prime \in \{1,...,N\}$;\\
    	(1-\lambda_n)^N, & if $s^\prime=s+N$; \\
    	0, & otherwise,
    	\IEEEstrut
    	\end{IEEEeqnarraybox}
    	\right.
    \end{IEEEeqnarray}
    by noticing the fact that the AoI increases by $N$ during successive RR-ONE scheduling and then is updated to a certain value given by \eqref{a_tau}. If no packet is present at the terminal queue, then the AoI is updated to $s^\prime=s+N$. In the following, the resultant steady-state stationary stationary distribution of the Markov chain, denoted by $\mu_n(s)$ where $s=1,2,...$ denotes state value, is derived. It follows that 
    \begin{equation}
    \mu_n(j) = \sum_{i \in \{1,2,...\}} \mu_n(i) p_{ij}.   
    \end{equation}
    Since 
    \begin{equation}
        p_{ij} = \lambda_n (1-\lambda_n)^{j-1},\,\forall i=\{1,2,...\},j \in \{1,...,N\},
    \end{equation}
    the steady-state stationary distribution for the first $N$ states is 
    \begin{equation}
    \mu_n(j) = \lambda (1-\lambda_n)^{j-1},\,\forall j \in \{1,...,N\}.
    \end{equation}
    For the $j$-th state with $j>N$, its previous state must be state $j-N$, and hence
    \begin{IEEEeqnarray}{rCl}
    &&\mu_n(j) = \mu_n(j-N)(1-\lambda_n)^N=...=\mu_n(l)(1-\lambda_n)^{mN} \nonumber\\
    &=& \lambda_n (1-\lambda_n)^{j-1},\,\forall j = mN + l,
    \end{IEEEeqnarray}
    wherein $m$ and $l$ are integers larger or equal to one. Therefore, it has been shown that the steady-state stationary distribution of the Markov chain is a geometric distribution with parameter $\lambda_n$. 
    
    After addressing the steady state of the embedded Markov chains, we are ready to derive the steady-state stationary distribution of the AoI evolution process $\left\{Y_n(t),\,t=1,2...\right\}$ at all time slots. Notice that it is a Markov renewal process with fixed renewal interval $N$ based on RR-ONE. Moreover, the state evolution between renewal (scheduling) is deterministic; the AoI increases by one every time slot. Therefore, the steady-state stationary distribution of $\left\{Y_n(t),\,t=1,2...\right\}$ is derived as 
    \begin{equation}
    \label{o4}
        \mu_n(j) = \sum_{m=1}^{\min[j,N]} q_{n,m,j} = \sum_{m=1}^{\min[j,N]} \frac{1}{N}\lambda_n (1-\lambda_n)^{j-m},
    \end{equation}
    where $q_{n,m,j}$ is the fraction of time that the AoI state transits to state-$(j-m+1)$ at the scheduled time slot and then reaches state-$j$ before the next scheduling. Therefore 
    \begin{equation}
    j-(j-m+1) \le N-1,
    \end{equation}
    such that state-$j$ is within reach, and hence $m \ge N$; this, combining with $m \le j$, explains the minimum operation in \eqref{o4}. The steady-state stationary distribution is hence given in \eqref{eq_dist} after some mathematical manipulations. The time-average AoI can be directly derived from this distribution, i.e., 
    \begin{IEEEeqnarray}{rCl}
    \label{rr_aoi}
        \bar{h}_{\mathsf{RR}}^{(\infty,N)} &=& \frac{1}{N} \sum_{n=1}^N \sum_{j=1}^\infty \mu_n(j) j  = \frac{1}{N} \sum_{n=1}^N \frac{1}{\lambda_n} + \frac{N-1}{2}.
    \end{IEEEeqnarray}
    This coincides with \eqref{eq_1}.    
    \begin{algorithm}[!t]
    	\caption{DRR}
    	\label{alg:rr}
    		\label{alg:init}
    		Initialization: \\
    		(Terminals) Set $\gamma_n = n \textrm{ mod } N$, $w_n=N$.\\
    		(BS) Set $N$ to the number of initial terminals and $W = 1$. \\
    		(Newly appeared terminals) Set $\gamma_n = 0$, $w_n=1$.\\
    		At each terminal: \\
    		\label{alg:start}
    		\If{Packet arrives}{Replace the current queued packet with the new one.}
    		\If{$\gamma_n=0$}{Transmit the queued packet (a blank packet if none).\\Set $\gamma_n \leftarrow w_n$.}
    		\Else{Keep silent in this time slot.}
    		At the BS, after receiving updates:\\
    		\If{Successful update or a blank update }{Send an ACK.}
    		\ElseIf{Receive nothing}{Send a NACK.}
    		\Else{(Collision) Set $N \leftarrow N+1$. Send a packet indicating a collision, containing information of $N$ and $W$.}
    		Set $W \leftarrow (W-1)$ mod $N$.\\
    		After receiving feedback at each terminal:\\
    		\If{Collision}{
    		\If{Terminal is newly appeared}{$w_n \leftarrow N$, $\gamma_n \leftarrow W$. $\gamma_n \leftarrow \gamma_n - 1$. }
    		\Else{$w_n \leftarrow N$, $\gamma_n \leftarrow N$. $\gamma_n \leftarrow \gamma_n - 1$. }
    		}
    		\ElseIf{NACK}{$w_n \leftarrow w_n-1$. $\gamma_n \leftarrow \gamma_n - 1$. }
    		\Else{ACK received. $\gamma_n \leftarrow \gamma_n - 1$. }
    		Return to Step \ref{alg:start}.
    \end{algorithm}
    \section{Decentralized RR-ONE Algorithm}
    \label{sec_dp}
    A fully decentralized RR-ONE-based scheduling algorithm (DRR) is proposed in this section. The proposed algorithm is described in Algorithm \ref{alg:rr}, where we assume dynamic terminal appearance. However, we assume at each time slot at most one event can happen: a terminal appears or disappears.
    
    The essence of DRR can be summarized as follows. The system always runs a round-robin status update protocol by assigning each terminal a unique time slot ($\gamma_n$) to update and rotating among terminals. Without new terminal appearances or disappearances, the system runs collision-free and every time slot is utilized. The BS feeds an ACK back in this case. Terminals dynamically appear at or disappear from the system. With a new terminal appearance, it updates immediately by design, and causing a collision inevitably since previously all time slots are utilized. The BS then feeds back a common message indicating a collision, with also the information about the current number of terminals (including the newly appeared), i.e., $N$, and the current spot of rotation, i.e., $W$. The new terminal will occupy the newly created time slot at the end of the rotation and the collided time slot still belongs to its original owner, and therefore no collision will happen between the two terminals in the next round. With a terminal (T$0$) disappearance, the BS receives nothing in its time slot. Note that a terminal that does not have any packet to update when it is scheduled would transmit a special blank packet such that the BS can distinguish between terminal disappearance and no packet to update. When T$0$ disappears, the BS then feeds back a NACK. Every terminal in the system then subtracts its record of the number of terminals by one and the next-in-line terminal (T$1$) will occupy T$0$'s spot in the next round. 
    
    \section{Proof of Theorem \ref{thm2}: Optimal AIR Policy without Packet Management for Peak-AoI Minimization}
    \label{sec_nopm}
    Without packet management, the status packets are assumed to be queued with infinite buffers at terminals before transmissions and the service is based on FCFS discipline in this paper. Therefore, all generated packets are eventually transmitted and updated to the BS. This assumption is reasonable in scenarios where not only the latest status is important, but also the BS is interested in keeping all the status data, and their status evolution processes, for analytic purposes. Another reasonable scenario is that terminals cannot apply packet management due to hardware limitations. 
    
    Note that the ASTA property does not apply in this case, due to the fact that the age of the head-of-line packet at each terminal changes to the next-in-line packet's age when the terminal is scheduled. Therefore, the independence property does not hold in Lemma \ref{lm_asta}; this presents a difficulty in analyzing the time-average AoI. Hence, we resolve to the PAoI metric in this section, which is defined as the AoI just before each update, i.e., the peaks of age evolution in Fig. \ref{fig_aoi}. The time-average PAoI is defined as
    \begin{equation}
    \label{paoi}
        \bar{\Lambda}_{\pi}^{(\infty,N)}  \triangleq \lim_{T \to \infty}  \frac{1}{N} \sum_{n=1}^N \frac{1}{K^{(T)}_n} \sum_{k_n=1}^{K^{(T)}_n}  \Lambda_{n,\pi}(k_n), 
    \end{equation}
    where $K^{(T)}_n$ denotes the number of status updates of terminal-$n$ up to time $T$, and $\Lambda_{n,\pi}(k_n)$ denotes the PAoI before the $k_n$-th update of terminal $n$ under policy $\pi$. The PAoI metric is used in many previous papers \cite{huang15,costa16} which show that the time-average PAoI can well approximates the AoI when the arrival processes have small second moments \cite[Lemma 1]{huang15}. The following optimization problem is considered.
    \begin{flalign}
    \label{p1}
    \textbf{P1:}&&\mathop{\textrm{minimize}}\limits_{\pi \in \mathcal{U}}  \,\,& \bar{\Lambda}_{\pi}^{(\infty,N)} &&\nonumber\\
    &&\textrm{s.t.,}\,\, &  \textrm{all queues are stable}, &&
    \end{flalign}
    where $\mathcal{U}$ denotes the set of all policies. The queue stability constraint is added for following reasons: first, queue stability is sufficient to ensure that when $T \to \infty$, $K^{(T)}_n \to \infty,\,\forall n = 1,...,N$, i.e., every terminal keeps updating its status; on the other hand, it is straightforward that a necessary condition for age stability is queue stability, by observing that when a queue is not, e.g., strongly stable \cite{neely10}, the corresponding AoI also goes to infinity since each queued packet increases the AoI at least by one. Note that without the queue stability constraint, optimizing the PAoI is problematic since totally abandoning some terminals may obtain better performance because the PAoI is only counted when updated. The queuing model for each terminal is represented by $M/G/1$ with service process $\mathcal{S}_n$ resulting from a policy $\pi$. Assume the service time of $\mathcal{S}_n$ has finite first- and second-moments which are denoted by $m_n$ and $v_n$, respectively. Based on the P-K formula \cite{bert92}, the time-average PAoI for terminal-$n$ is
    \begin{IEEEeqnarray}{rCl}
        \bar{\Lambda}_{n,\pi}^{(\infty)} \overset{(a)}{=} \mathbb{E}[I_n + J_n] &=& \frac{1}{\lambda_n} + m_n + \frac{\lambda_n v_n}{2(1-\lambda_n m_n)} \nonumber\\
        &\overset{(b)}{\ge}& \frac{1}{\lambda_n} + \frac{1}{2}\left(\frac{1}{\beta_n-\lambda_n}+\frac{1}{\beta_n}\right),
    \end{IEEEeqnarray}
    where $I_n$ and $J_n$ denote the inter-arrival time and the sojourn time of packets in queue-$n$, respectively. The equality in $(a)$ is based on the definition of PAoI \cite{huang15}. The mean service rate $\mathcal{S}_n$ is denoted by $\beta_n$. The last inequality $(b)$ is based on the fact that $v_n \ge m_n^2$, which means that it is best to have a deterministic service process to minimize the mean delay; the same conclusion in fact hold for $GI/G/1$ queuing models \cite{whitt84}. The PAoI minimization problem of \textbf{P1} is thus transformed to \textbf{P2} presented in Section \ref{sec_sm}. Denote the objective function as $f(\bm{\beta}) \triangleq \frac{1}{N} \sum_{n=1}^N f_n(\beta_n)$ where
    \begin{equation}
        f_n(\beta_n) \triangleq \frac{1}{\lambda_n} + \frac{1}{2}\left(\frac{1}{\beta_n-\lambda_n}+\frac{1}{\beta_n}\right).
    \end{equation}
    Then $f(\bm{\beta})$ is a strictly convex function since 
    \begin{equation}
        \frac{\textrm{d}^2 f_n(\beta_n)}{\textrm{d}^2 \beta_n} = \frac{1}{(\beta_n-\lambda_n)^3} + \frac{1}{\beta_n^3} > 0,\,\forall n=1,...,N.
    \end{equation}
    Combining with the fact that the constraints in \textbf{P2} are all linear constraints, therefore, \textbf{P2} is a convex optimization \cite{boyd09} problem which can be solved efficiently. After obtaining the optimum service rates $\bm{\beta}_\textrm{opt}$, the optimal policy is then to serve every terminal regularly with fixed intervals given by $\bm{\beta}_\textrm{opt}$. Note that this may not be possible with given, possibly irrational number, $\bm{\beta}_\textrm{opt}$. 
    
    Unfortunately, it seems that closed-form optimal solutions are elusive for \textbf{P2}. We then find a reasonably good closed-form solution in Lemma \ref{lm_ub} (cf. proof in Appendix \ref{app_lemmas}) to shed light on some insights.    
    \begin{lemma}
    \label{lm_ub}
    Define the traffic intensity of terminal $n$ as $\rho_n \triangleq \frac{\lambda_n}{\beta_n}$. The minimum time-average PAoI is upper bounded by 
    \begin{equation}
    \label{ub}
        f(\bm{\beta}_\textrm{ub}) = \frac{1}{N} \sum_{n=1}^N \left[\frac{1}{\lambda_n} + \frac{1}{2}\left(\frac{N}{\epsilon}+\frac{1}{\frac{\epsilon}{N} + \lambda_n}\right)\right],
    \end{equation}
    where $\epsilon \triangleq 1-\sum_{n=1}^N \lambda_n$. This bound is achieved by 
    \begin{equation}
    \label{ub_app}
    \beta_{n,\textrm{ub}} = \frac{\epsilon}{N} + \lambda_n,\, \forall n=1,...,N.
    \end{equation}
    With heavy traffic, i.e., $\rho_n \to 1$, $\forall n=1,...,N$, this solution is asymptotically optimal. $\hfill\square$
    \end{lemma}
    
    \begin{remark}
    By comparing the results in Theorem \ref{thm1} and Lemma \ref{lm_ub} in the asymptotic regime, some insights can be obtained. First, in the massive IoT regime, the optimum scaling factor without packet management is
    \begin{equation}
    \lim_{N \to \infty} \frac{f(\bm{\beta}_\textrm{ub})}{N} = \frac{1}{2 \epsilon}.
    \end{equation}
    Compared with the optimum scaling factor with packet management, i.e., $0.5$, the packet waiting delay in the queue introduces a multiplicative loss of $1/\epsilon$ which is related to traffic intensity. 
    
    The optimum scaling laws with diminishing arrival rates are actually identical with and without packet management. This is intuitive since the bottleneck is arrival status update packets in this regime, and packets are rarely queued with or without packet management. 
    \end{remark}
    
    \section{Simulation Results}
    \label{sec_sr}
    In this section, computer simulation based experiments are conducted to evaluate the AoI performance of scheduling policies. This enables us to obtain the optimum performance numerically by formulating the problem as an Markov decision process (MDP) and solving it by relative value iterations for average cost function \cite{bersk}. Note that, similar with most practical applications, the MDP based approach suffers from the curse of dimensionality and hence only small-scale problems can be solved thereby. Specifically, it is observed that the state space size grows exponentially with the number of terminals, and hence the scalability is significantly limited. Nevertheless, we obtain the minimum time-average AoI of a $2$-terminal case and compare its performance with RR-ONE. An finite-state approximation is made for the MDP which originally has infinite state space; note that the AoI can grow to infinity. However, the optimality is intact by arguing that the optimal policy given by solving the finite state MDP does not allow the AoI grows to our prescribed AoI limit. The state space of the MDP is defined as 
    \begin{equation}
        (h_1,\,a_1) \times (h_2,\,a_2),
    \end{equation}
    where $h_i$ denotes the AoI at the BS for terminal $i$, $a_i$ denotes the age of the packet at terminal $i$ (assuming the terminal adopts the one-packet buffer packet management policy), and $1 \le h_i,a_i \le h_\textrm{max}$, $i=1,2$, where $h_\textrm{max}$ is the age limit. The transition probability matrix follows straightforwardly; it is omitted here for brevity, along with the relative value iteration procedure which is well known. The performance of RR-ONE is obtained by running RR-ONE for $10^5$ time slots and calculating the time-average AoI. In addition, we also simulate UN-ONE which schedules a terminal uniformly random at each time slot, and an age-greedy policy which chooses the terminal with the largest AoI. In fact, UN-ONE can be regarded as the CSMA scheme which is shown to be optimal to maximize throughput with greedy sources. The age-greedy policy is found optimal without considering random packet arrivals \cite{kadota16}. 
    \begin{figure}[!t]
    \centering
    \includegraphics[width=0.45\textwidth]{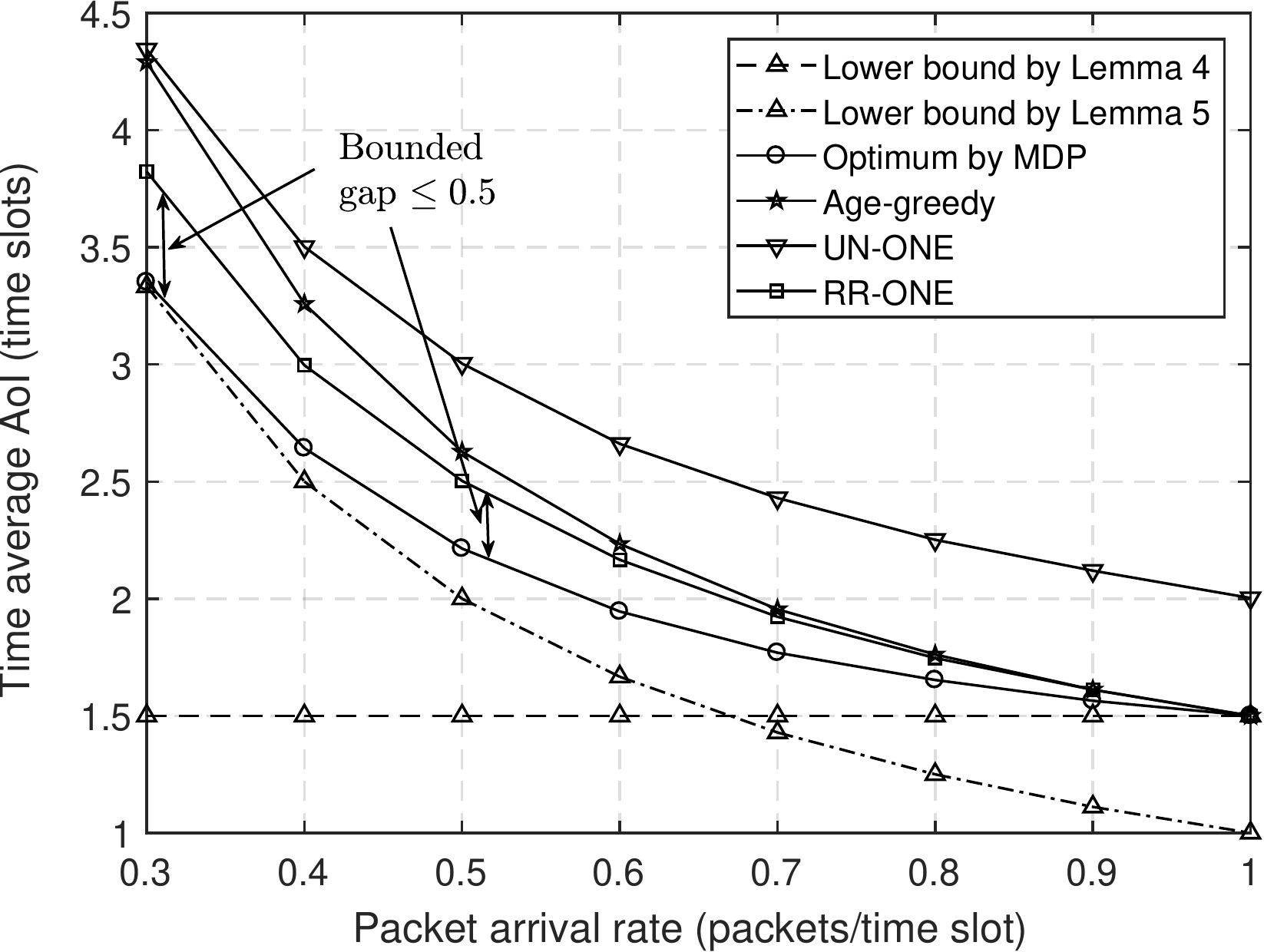}
    \caption{Performance comparisons among lower bounds in Lemma \ref{lma0} and \ref{lma1}, MDP-based optimum, UN-ONE, age-greedy policy and RR-ONE. There are $2$ terminals with identical packet arrival rate (x-axis).}
    \label{fig_mdp}
    \end{figure}
    \begin{figure}[!t]
    \centering
    \includegraphics[width=0.45\textwidth]{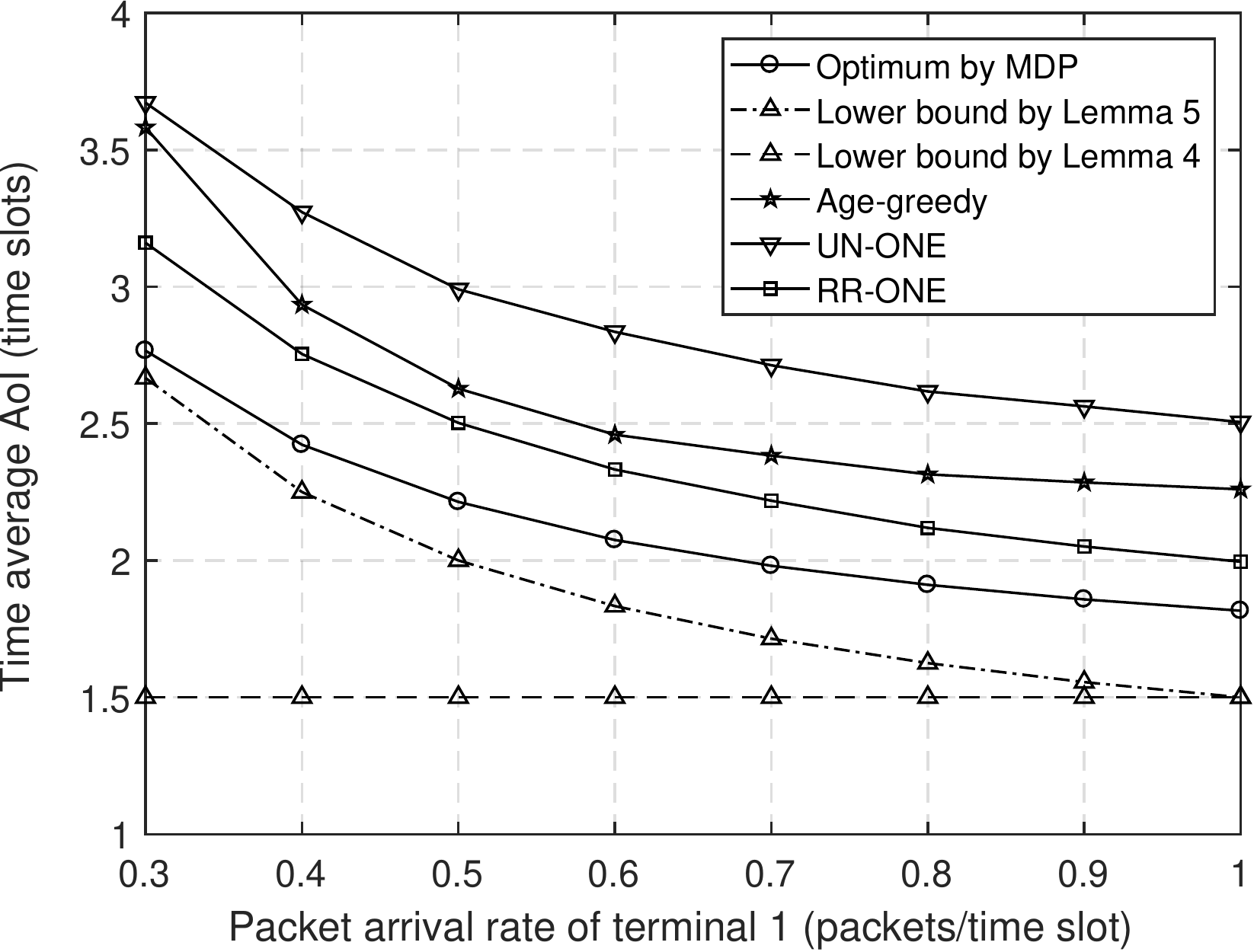}
    \caption{Performance comparisons with heterogeneous arrival rates. There are $2$ terminals; $\lambda_2=0.5$ and $\lambda_1$ is shown as the x-axis.}
    \label{fig_mdp_1}
    \end{figure}
    
    It is observed from Fig. \ref{fig_mdp} that the performance gap between RR-ONE and the optimum given by numerically solving the MDP is larger with lower packet arrival rates; on the other hand, RR-ONE achieves the optimum when $\lambda$ approaches one based on Remark \ref{one_remark}. Since RR-ONE is proved optimal among AIR policies, the optimal scheduling policy with low packet arrival rates must be a non-AIR policy. Specifically, the following intuition explains this. Suppose that the probability of both terminals having arrival packets in the same time slot is negligible when arrival rates are sufficiently low; then the optimal policy is immediately obvious that it should schedule the terminal with a packet arrival at each time slot; note that this policy is not an AIR policy because the scheduling decision depends on packet arrivals which violates Condition $1$ of AIR policy definition. The optimum performance in this case is also obvious; it should be the same with what is shown in Lemma \ref{lma1}, i.e., completely determined by the inter-arrival time of packets and this can be observed from Fig \ref{fig_mdp}. Nevertheless, it is noted that the performance gap is bounded (within $0.5$ time slots) in this $2$-terminal case by observing the RR-ONE performance and performance bound in Lemma \ref{lma1}. 
    
    Fig. \ref{fig_mdp_1} shows the performance with heterogeneous packet arrival rates for terminals; the arrival rate $\lambda_1$ is fixed to $0.5$ packets per time slot and $\lambda_2$ varies from $0.3$ to $1$. The bounded gap is still within $0.5$ time slots, despite the fact that RR-ONE disregards the heterogeneity of arrival rates completely. Nonetheless, it is observable that the gap between optimum and RR-ONE is larger compared with Fig. \ref{fig_mdp}. Moreover, the gap between RR-ONE and the age-greedy policy is relatively larger when the arrival rates difference between terminals increases, showing that the age-greedy is more sensitive to arrival statistics heterogeneity.
    \begin{figure}[!t]
    \centering
    \includegraphics[width=0.45\textwidth]{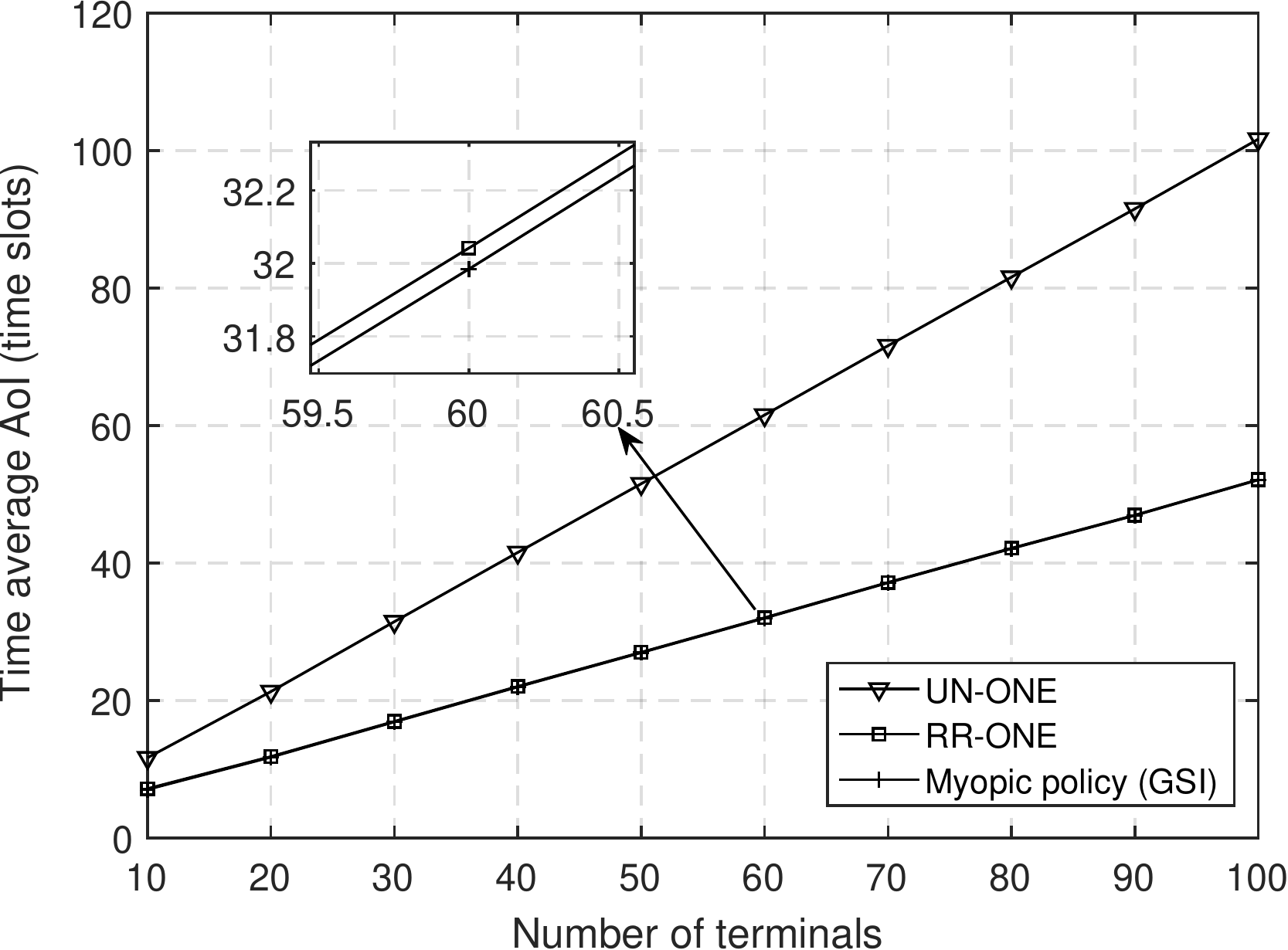}
    \caption{Performance comparisons with a large number of terminals; the arrival rates are uniformly randomly generated from $[0,1]$.}
    \label{fig_rr_N}
    \end{figure}
    \begin{figure}[!t]
    \centering
    \includegraphics[width=0.45\textwidth]{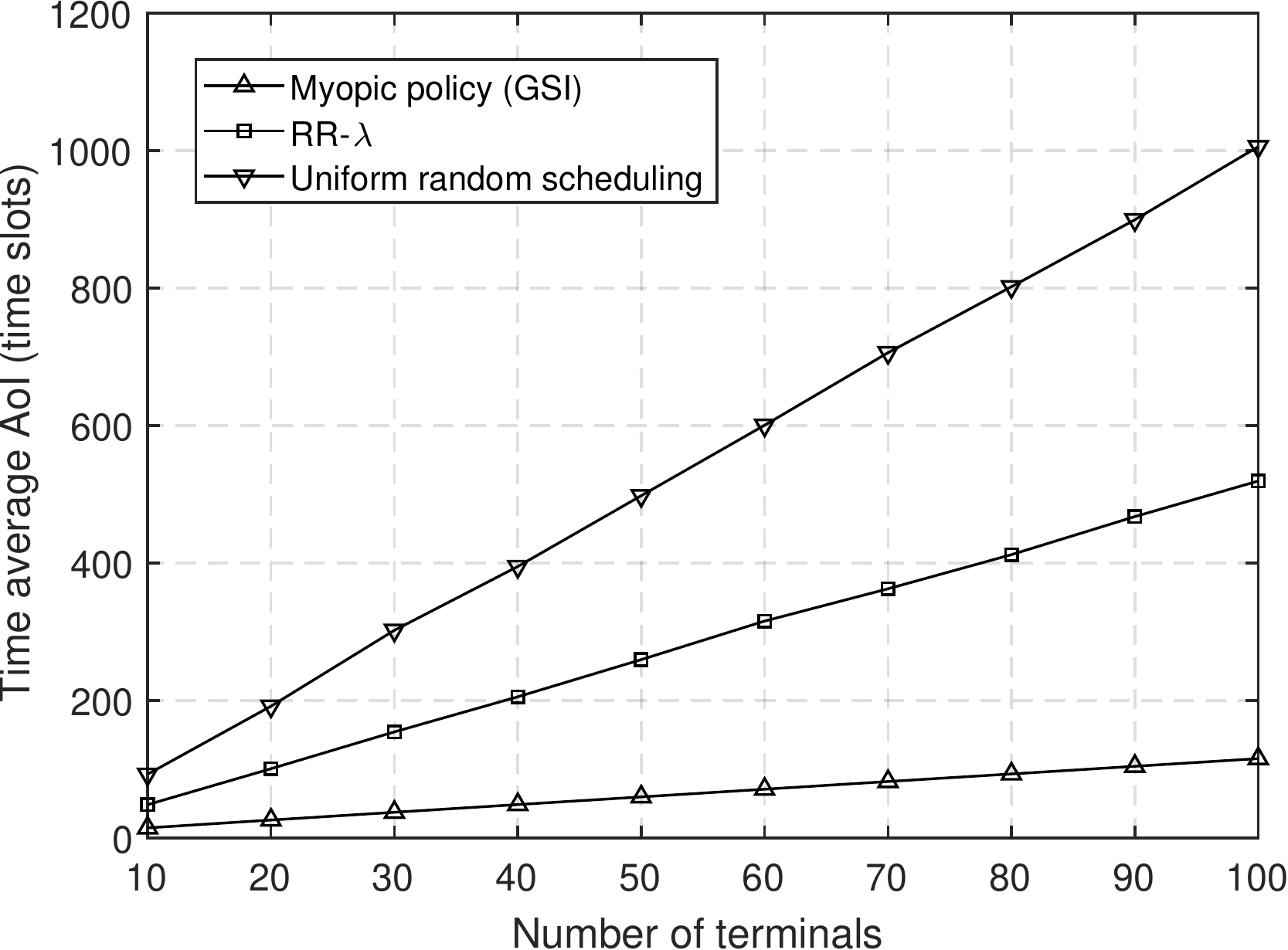}
    \caption{Performance comparisons with a large number of terminals and no packet management; the arrival rates are identical for all terminals, and $\epsilon=0.1$ in Lemma \ref{lm_ub}.}
    \label{fig_un_N}
    \end{figure}
    
    We increase the number of terminals and enter the massive IoT regime in Fig. \ref{fig_rr_N}. The MDP-based optimal solution is computationally intractable in this regime and hence we adopt the myopic policy with GSI as an approximation of the optimum. The myopic policy with GSI leverages all the global information (though no future knowledge) to make a scheduling decision that minimizes the one-step expected AoI cost in the MDP formulation; by comparing it with RR-ONE helps us to understand how much GSI benefits RR-ONE. Based on Fig. \ref{fig_rr_N}, it is shown that the myopic policy with GSI outperforms RR-ONE only slightly, due to the reason that the packet management of using one-packet buffers eliminates most of the randomness of packet arrivals; most packets are dropped by packet management due to staleness and hence their randomness has no effect. The performance of UN-ONE is also shown; it has been proved in Corollary \ref{coro_un} that its linear scaling factor is (at least) $1$ compared with $1/2$ for RR-ONE; this can be observed in the figure.
    
    In Fig.\ref{fig_un_N}, policies without packet management are simulated. The arrival packets are queued at terminals and all arrived packets have to be delivered to the BS. The optimal AIR policy RR-$\bm{\lambda}$ in this case is compared with random scheduling and myopic policy with GSI in the figure. It is observed that, without packet management, the gap between the myopic policy with GSI and the optimal AIR policy RR-$\bm{\lambda}$ is much larger compared with that (between myopic policy with GSI and RR-ONE) in Fig. \ref{fig_rr_N}; in fact different asymptotic scaling factors are found in Fig. \ref{fig_un_N}. The intuition behind this is that the packet arrival randomness has much larger impact if all packets are queued; therefore the GSI in this case is more valuable compared with the case with packet management whereby most randomness is eliminated. The performance advantage of RR-$\bm{\lambda}$ over uniformly random scheduling is still evident. Note that by setting all arrival rates to be identical and the associated parameters in Fig. \ref{fig_un_N}, all three policies can stabilize the system; it is pointless to evaluate a policy that cannot stabilize the system since it would lead to infinite AoI. 
    
    \section{Conclusions and Discussions}
    \label{sec_cl}
    In this paper, it is found that with a number of terminals sharing a common wireless uplink based on a collision model and random packet arrivals at each terminal, the optimal AIR policy to minimize the time-average AoI is RR-ONE, i.e., scheduling terminals in a round-robin fashion and each terminal only retains the most up-to-date packet. In the asymptotic regime where the number of terminals is large, the optimum (among all policies) time-average AoI is proved to scale linearly with the number of terminals with the optimum scaling factor of $\frac{1}{2}$, and RR-ONE achieves the optimum asymptotically. The steady-state stationary distribution of the AoI of each terminal under RR-ONE is also derived. In addition to establishing the optimality, we propose a full-fledged decentralized algorithm to implement RR-ONE which accounts for dynamic terminal appearances. Considering the scenario where the entire procedure of status variation is requested, i.e., packets cannot be dropped, the optimal AIR policy without packet management, namely RR-$\bm{\lambda}$, is also derived; it schedules terminals based on deterministic scheduling intervals $m_n$, $n=1,...,N$, with $m_n$ given by the solution of a convex optimization problem. 
    
    Both RR-ONE and RR-$\bm{\lambda}$ are suitable for decentralized implementation since they require minimal information for scheduling decisions. Based on numerical simulations, it is observed that extra global information hardly benefits RR-ONE; on the other hand, it benefits RR-$\bm{\lambda}$ much more evidently (also linear scaling of AoI with the number of terminals but with a smaller scaling factor). This stems from the fact that the aggressive packet management by RR-ONE significantly eliminates the randomness of packet arrivals since a considerable share of packets are dropped due to staleness; however this randomness accumulates and has a more noticeable effect without packet management by RR-$\bm{\lambda}$. The extra global information for scheduling is mainly to deal with the arrival randomness, and therefore it is less valuable to RR-ONE compared with RR-$\bm{\lambda}$. The packet management of RR-ONE essentially ``hardens'' the packet arrival processes, i.e., making them more deterministic, by observing that the performance of RR-ONE is analogous with that of deterministic packet arrivals (comparing Lemma \ref{lma0}, Lemma \ref{lma1} and Theorem \ref{thm_air}).  
    
    \appendices
    \section{Myopic Policy with GSI is not Optimal with Finite Horizon}
    \label{app1}
    In this section, we assume that the BS knows the current ages of packets of all terminals and the arrival distributions, i.e., $\lambda_i$, $i \in \{1,...,N\}$ to make a centralized scheduling decision. In this case, it will be shown the myopic policy is not optimal by constructing a counter example. Consider a case with a time horizon $T=2$, and the number of terminals is $N=2$. The arrival rates for terminal $1$ and $2$ are $\lambda_1 = \delta$ and $\lambda_2 = 1-\delta$, respectively, and $0<\delta<1$. The initial ages are set as $h_i(0)$, $i=1,2$, and the initial age gains of the packets in the queues are $g_i(0)$, $i=1,2$. Assuming the following conditions are met:
    \begin{equation}
    \frac{g_2(0)}{2} < g_1(0) < g_2(0),
    \end{equation}
    \begin{equation}
    h_2(0) - g_2(0)  > h_1(0),
    \end{equation}
    then the expected average age under the myopic policy, which always schedules the terminal with the largest $g_n(t)$ at time $t$, is shown here
    \begin{IEEEeqnarray}{rCl}
    \label{ex1}
    &&\frac{1}{4}\left(h_{1,\textrm{mo}}(1)+h_{1,\textrm{mo}}(2)+h_{2,\textrm{mo}}(1)+h_{2,\textrm{mo}}(2)\right) \nonumber\\
    &=& C - \frac{1}{4}\left(g_2(0) + \delta(1-\delta)(h_2(0)+1) \right. \nonumber\\
    && + \delta^2 (g_2(0) + h_1(0)+1) + \delta(1-\delta)(g_1(0)+g_2(0)) \nonumber\\
    && \left.+ (1-\delta)^2(h_2(0)+1)  \right) \nonumber\\
    &=& C - \frac{1}{4}\left(g_2(0) + h_2(0)+1 + \delta (g_1(0)-g_2(0)) \right. \nonumber\\
    &&  \left.+ \delta^2 (h_1(0) + 1 - g_1(0))\right),
    \end{IEEEeqnarray}
    and the expected average age under the optimal policy, which is not hard to figure out in this simple case, is 
    \begin{IEEEeqnarray}{rCl}
    \label{ex2}
    &&\frac{1}{4}\left(h_{1,*}(1)+h_{1,*}(2)+h_{2,*}(1)+h_{2,*}(2)\right) \nonumber\\
    &=& C - \frac{1}{4}\left(g_1(0) + \delta(1-\delta)(g_1(0)+h_2(0)+1) \right.\nonumber\\
    && + \delta^2 (g_1(0) + g_2(0)) + \delta(1-\delta)(g_1(0)+g_2(0)) \nonumber\\
    && \left.+ (1-\delta)^2(g_1(0)+h_2(0)+1)  \right) \nonumber\\
    &=& C - \frac{1}{4}\left(2g_1(0) + h_2(0)+1 - \delta (h_2(0) + 1 - g_2(0))\right),\nonumber\\
    \end{IEEEeqnarray}
    where 
    \begin{equation}
    C=\frac{1}{2}\left(h_{1}(0) + h_{2}(0)\right) + 1.5.
    \end{equation}
    By setting the parameter $\delta$ sufficiently small, we can see that the average age of the myopic policy is strictly smaller than the optimal policy which schedules terminal $1$ at the first time slot although the current myopic choice is terminal $2$.
    
    \section{Proof of Lemmas}
    \label{app_lemmas}        
    \begin{IEEEproof}[Proof of Lemma \ref{lm_0}]
    \label{proof_lm0}
    It is straightforward that the performance of a non-WCNC policy is stochastically dominated by a WCNC policy which is identical to the non-WCNC policy, except that the WCNC policy schedules an arbitrary terminal (resp. schedules one of the terminals) when the non-WCNC policy is idle (resp. schedules multiple terminals resulting in collisions). 
    \end{IEEEproof}
    
    \begin{IEEEproof}[Proof of Lemma \ref{lm_1}]
    \label{proof_lm1}
    At each time slot, the probability of a packet arrival is $\lambda$ and therein the age decreases to one; otherwise, with probability $1-\lambda$, the age increases by one. The Markovian property holds obviously. The steady-state stationary distribution is hence a geometric distribution with parameter $\lambda$.
    \end{IEEEproof}

    \begin{IEEEproof}[Proof of Lemma \ref{lma0}]
    Consider a system $\mathsf{A0}$ wherein the AoI after each update is fixed to one, instead of determined by the random arrival packets. Then for any scheduling policy $\pi$, we have
    \begin{equation}
    \label{o2}
        \bar{h}_{\pi}^{(T,N)} \ge \bar{h}_{\pi,\mathsf{A0}}^{(T,N)}, 
    \end{equation}
    where $\bar{h}_{\pi,\mathsf{A0}}^{(T,N)}$ denotes the time-average AoI of policy $\pi$ in system $\mathsf{A0}$. The proof for \eqref{o2} is simply based on stochastic dominance, and currently omitted for brevity. By \cite[Theorem 1]{kadota16}, the optimal policy in $\mathsf{A0}$ is the age-greedy policy by noticing that no asymmetric transmission failure or weights are considered in this paper. Therefore it is straightforward to derive the optimum time-average AoI in $\mathsf{A0}$ since no randomness exists. The optimum AoI in $\mathsf{A0}$ is therefore
    \begin{equation}
        \bar{h}_{\textrm{opt},\mathsf{A0}}^{(\infty,N)} = \frac{N+1}{2}, 
    \end{equation}
    
    It is sufficient to consider the optimal scheduling scheme in $\mathsf{A0}$ to obtain a AoI lower bound since all scheduling schemes perform better in $\mathsf{A0}$ compared with a system with random packet arrivals. The conclusion follows immediately. 
    \end{IEEEproof}

    \begin{IEEEproof}[Proof of Lemma \ref{lma1}]
    To prove this result, we consider a collision-free system $\mathsf{A1}$ wherein the uplink transmissions can be multiplexed, i.e., an arbitrary number of terminals can update successfully in the same time slot. It is obvious that, similar with Lemma \ref{lma0}, for any policy $\pi$,
    \begin{equation}
    \label{o3}
        \bar{h}_{\pi}^{(T,N)} \ge \bar{h}_{\pi,\mathsf{A1}}^{(T,N)}.
    \end{equation}
    It is also obvious that the time-average AoI in system $\mathsf{A1}$ is the time-average packet inter-arrival time at terminal queues. Therefore, the conclusion follows immediately.
    \end{IEEEproof}

    \begin{IEEEproof}[Proof of Lemma \ref{lm_ub}]
    Since \textbf{P2} is minimizing the PAoI over all $\beta_n$, by setting $\beta_n = \beta_{n,\textrm{ub}}$ we can obtain the upper bound in Lemma \ref{lm_ub}. In the heavy traffic regime, 
    \begin{IEEEeqnarray}{rCl}
        f(\bm{\beta}) &=& \frac{1}{N} \sum_{n=1}^N \frac{1}{\lambda_n} + \frac{\rho_n}{2\lambda_n}\left(1+\frac{1}{1-\rho_n}\right) \nonumber\\
        &\overset{\rho_n \to 1}{\longrightarrow}& \frac{1}{N} \sum_{n=1}^N \frac{1}{2 \lambda_n (1-\rho_n)}.
    \end{IEEEeqnarray}
    The optimum solution in the heavy traffic regime is therefore $\beta_{n,\textrm{ub}}$, $\forall n=1,...,N$ by applying the arithmetic-harmonic-mean inequality.
    \end{IEEEproof}
    
    \bibliographystyle{ieeetr}
    \bibliography{aoi}

\begin{thebibliography}{10}

\bibitem{jiang18_isit}
Z.~Jiang, B.~Krishnamachari, X.~Zheng, S.~Zhou, and Z.~Niu, ``Decentralized
  status update for age-of-information optimization in wireless multiaccess
  channels,'' in {\em submission to IEEE Int'l Symp. Info. Theory}, 2018.

\bibitem{ericsson_iot}
Ericsson, ``Cellular networks for massive {IoT},'' {\em tech. rep.}, Jan 2016.

\bibitem{oss14}
A.~Osseiran, F.~Boccardi, V.~Braun, K.~Kusume, P.~Marsch, M.~Maternia,
  O.~Queseth, M.~Schellmann, H.~Schotten, H.~Taoka, H.~Tullberg, M.~A.
  Uusitalo, B.~Timus, and M.~Fallgren, ``Scenarios for {5G} mobile and wireless
  communications: The vision of the {METIS} project,'' {\em IEEE Commun. Mag.},
  vol.~52, pp.~26--35, May 2014.

\bibitem{fet14}
G.~P. Fettweis, ``The {Tactile Internet}: Applications and challenges,'' {\em
  IEEE Veh. Tech. Mag.}, vol.~9, pp.~64--70, Mar 2014.

\bibitem{kaul11}
S.~Kaul, M.~Gruteser, V.~Rai, and J.~Kenney, ``Minimizing age of information in
  vehicular networks,'' in {\em IEEE SECON}, pp.~350--358, Jun 2011.

\bibitem{kaul12}
S.~Kaul, R.~Yates, and M.~Gruteser, ``Real-time status: How often should one
  update?,'' in {\em IEEE INFOCOM}, pp.~2731--2735, Mar 2012.

\bibitem{yates12}
R.~D. Yates and S.~Kaul, ``Real-time status updating: Multiple sources,'' in
  {\em IEEE Int'l Symp. Info. Theory}, pp.~2666--2670, Jul 2012.

\bibitem{yates16}
R.~D. Yates and S.~K. Kaul, ``The age of information: Real-time status updating
  by multiple sources,'' {\em arXiv preprint arXiv:1608.08622}, 2016.

\bibitem{kadota16}
I.~Kadota, E.~Uysal-Biyikoglu, R.~Singh, and E.~Modiano, ``Minimizing the age
  of information in broadcast wireless networks,'' in {\em Annu. Allerton Conf.
  Commun., Control, Comput.}, pp.~844--851, Sep 2016.

\bibitem{he17}
Q.~He, D.~Yuan, and A.~Ephremides, ``Optimal link scheduling for age
  minimization in wireless systems,'' {\em IEEE Trans. Inform. Theory}, in
  press.

\bibitem{joo17}
C.~Joo and A.~Eryilmaz, ``Wireless scheduling for information freshness and
  synchrony: Drift-based design and heavy-traffic analysis,'' in {\em IEEE
  WiOpt}, pp.~1--8, May 2017.

\bibitem{jiang17_iotj}
Z.~Jiang, S.~Zhou, X.~Guo, and Z.~Niu, ``Task replication for
  deadline-constrained vehicular cloud computing: Optimal policy, performance
  analysis and implications on road traffic,'' {\em IEEE Internet Things J.},
  in press.

\bibitem{neely10}
M.~J. Neely, ``Stochastic network optimization with application to
  communication and queueing systems,'' {\em Synthesis Lectures on
  Communication Networks}, vol.~3, no.~1, pp.~1--211, 2010.

\bibitem{bia00}
G.~Bianchi, ``Performance analysis of the {IEEE} 802.11 distributed
  coordination function,'' {\em IEEE J. Select. Areas Commun.}, vol.~18,
  pp.~535--547, Mar 2000.

\bibitem{huang15}
L.~Huang and E.~Modiano, ``Optimizing age-of-information in a multi-class
  queueing system,'' in {\em IEEE Int'l Symp. Info. Theory}, pp.~1681--1685,
  Jun 2015.

\bibitem{bedewy16}
A.~M. Bedewy, Y.~Sun, and N.~B. Shroff, ``Optimizing data freshness,
  throughput, and delay in multi-server information-update systems,'' in {\em
  IEEE Int'l Symp. Info. Theory}, pp.~2569--2573, Jul 2016.

\bibitem{kaul12_ciss}
S.~K. Kaul, R.~D. Yates, and M.~Gruteser, ``Status updates through queues,'' in
  {\em Annual Conference on Information Sciences and Systems}, pp.~1--6, Mar
  2012.

\bibitem{costa16}
M.~Costa, M.~Codreanu, and A.~Ephremides, ``On the age of information in status
  update systems with packet management,'' {\em IEEE Trans. Inform. Theory},
  vol.~62, pp.~1897--1910, April 2016.

\bibitem{najm16}
E.~Najm and R.~Nasser, ``The age of information: The gamma awakening,'' in {\em
  IEEE Int'l Symp. Info. Theory}, pp.~2574--2578, 2016.

\bibitem{sun17}
Y.~Sun, E.~Uysal-Biyikoglu, R.~Yates, C.~E. Koksal, and N.~B. Shroff, ``Update
  or wait: How to keep your data fresh,'' in {\em IEEE INFOCOM}, pp.~1--9,
  April 2016.

\bibitem{hsu17}
Y.~P. Hsu, E.~Modiano, and L.~Duan, ``Age of information: Design and analysis
  of optimal scheduling algorithms,'' in {\em IEEE Int'l Symp. Info. Theory},
  pp.~561--565, Jun 2017.

\bibitem{whitt84}
W.~Whitt, ``Minimizing delays in the {GI/G/1} queue,'' {\em Operations
  Research}, vol.~32, no.~1, pp.~41--51, 1984.

\bibitem{Andrews14}
J.~Andrews, S.~Buzzi, W.~Choi, S.~Hanly, A.~Lozano, A.~Soong, and J.~Zhang,
  ``What will 5{G} be?,'' {\em IEEE J. Sel. Areas Commun.}, vol.~32,
  pp.~1065--1082, Jun 2014.

\bibitem{ahm09}
S.~H.~A. Ahmad, M.~Liu, T.~Javidi, Q.~Zhao, and B.~Krishnamachari, ``Optimality
  of myopic sensing in multichannel opportunistic access,'' {\em IEEE Trans.
  Inform. Theory}, vol.~55, pp.~4040--4050, Sep 2009.

\bibitem{ben95}
B.~Melamed and D.~Yao, ``The {ASTA} property,'' {\em Advances in Queueing:
  Theory, Methods and Open Problems}, pp.~195--224, 1995.

\bibitem{najm17}
E.~Najm, R.~Yates, and E.~Soljanin, ``Status updates through {M/G/1/1} queues
  with {HARQ},'' in {\em IEEE Int'l Symp. Info. Theory}, pp.~131--135, Jun
  2017.

\bibitem{cox67}
D.~R. Cox, {\em Renewal theory}, vol.~1.
\newblock Methuen London, 1967.

\bibitem{bert92}
D.~Bertsekas and R.~Gallager, {\em Data Networks}.
\newblock Prentice-Hall, Inc, 1992.

\bibitem{boyd09}
S.~Boyd and L.~Vandenberghe, {\em Convex Optimization}.
\newblock Cambridge university press, 2009.

\bibitem{bersk}
D.~P. Bertsekas, {\em Dynamic programming and optimal control}, vol.~1.
\newblock Athena scientific Belmont, MA, 1995.

\end{thebibliography}
    \end{document}